\documentclass[10pt,aps, superscriptaddress, notitlepage]{revtex4-2}
\usepackage{amsbsy}
\usepackage{amsmath,amssymb}
\usepackage[usenames]{color}
\usepackage[colorlinks,linkcolor=black,citecolor=blue]{hyperref}
\usepackage{graphicx}
\usepackage{xspace}
\usepackage{enumitem}
\usepackage{algpseudocode}
\usepackage[linesnumbered,ruled,vlined]{algorithm2e}
\usepackage{xcolor}
\usepackage{bm}
\usepackage{soul}
\usepackage{subfigure}
\usepackage{times}
\usepackage{physics}
\usepackage{tikz}
\usepackage{cleveref}
\usepackage{bbm}
\usetikzlibrary{positioning}
\usepackage[skins]{tcolorbox}
\usetikzlibrary{arrows.meta}
\usepackage{qcircuit}
\usepackage[absolute,overlay]{textpos}
\usetikzlibrary{shapes}
\newcommand\rd[0]{32mm} % radius for the circle placement
\usetikzlibrary{decorations.markings, decorations.text}           
\newcommand{\tikzbullet}[2]{%
 \begin{tikzpicture}[baseline=-0.5ex]
 \filldraw[draw=#1, fill=#2] (0,0) circle [radius=.25em];
 \end{tikzpicture}%
}
\definecolor{dbc}{rgb}{0.,0.4,.8}
\definecolor{ogc}{rgb}{0.4,.6,.5}

\newcommand{\uoh}{School of Physics, University of Hyderabad, India 500046}
\newcommand{\cqt}{Centre for Quantum Technologies, National University of Singapore, Singapore 117543}

\newcommand{\Qilimanjaro}{Qilimanjaro Quantum Tech, Barcelona 08007, Spain}
\newcommand{\TII}{Quantum Research Centre, Technology Innovation Institute, P.O.Box: 9639, Abu Dhabi, United Arab Emirates}
\newcommand{\IMRE}{Institute of Material Research and Engineering (IMRE), Agency for Science, Technology and Research (A*STAR), 2 Fusionopolis Way, Innovis \#08-03, Singapore 138634, Republic of Singapore}

\begin{document}

% \title{Trainability of a quantum-classical machine in the NISQ era}
\title{Practicality of training a quantum-classical machine in the NISQ era}
% \title{Experimental approach to NISQ training on a NISQ device}

\author{Tarun Dutta}
\email{tarunduttaz@gmail.com}
\address{\uoh}
\address{\cqt}

\author{Alex Jin}
\email{alex\_jin@u.nus.edu}
\address{\cqt}
\author{Clarence Liu Huihong}
% \email{}
\address{\cqt}
\author{José Ignacio Latorre}
\address{\cqt}
\address{\Qilimanjaro}
\address{\TII}

\author{Manas Mukherjee}
\email{manas.mukh@gmail.com}
\address{\cqt}
% \address{\maju}
\address{\IMRE}

\begin{abstract}
Advancements in classical computing have significantly enhanced machine learning applications, yet inherent limitations persist in terms of energy, resource and speed. Quantum machine learning algorithms offer a promising avenue to overcome these limitations but poses its own hurdles. This experimental study explores the limits of training a real experimental quantum classical hybrid system using supervised training protocols, on an ion trap platform. Challenges associated with ion trap-coupled classical processors are addressed, highlighting the $robustness$ of the genetic algorithm as a classical optimizer in navigating the noisy channels of NISQ-devices and the complex optimization landscape inherent in binary classification problems with many local minima. We intricately discuss why gradient-based optimizers may not be suitable in the NISQ era through thorough analysis. These findings contribute insights into the performance of quantum-classical hybrid systems, emphasizing the significance of efficient training strategies and hardware considerations for practical quantum machine learning applications. This work not only advances the understanding of hybrid quantum-classical systems but also underscores the potential impact on real-world challenges through the convergence of quantum and classical computing paradigms operating without the aid of classical simulators.
\end{abstract}
\maketitle
 \section{Introduction}\label{sec:intro}
The rapid progress in classical computing hardware has significantly advanced the field of machine learning, addressing a diverse range of real-world problems~\cite{Libbrecht2015, Vamathevan2019, Thirunavukarasu2023, Thiyagalingam2022, Xiao2024}. However, inherent limitations in speed and energy efficiency persist, as dictated by Moore’s and Koomey’s laws~\cite{658762, 5440129, 10.1117/12.209195} respectively. Quantum machine learning algorithms hold promise in overcoming these barriers, particularly as quantum hardware evolves to handle complex and practically useful problems~\cite{Biamonte2017, Xiao2023}. As quantum hardware scales up, understanding system-level implementation becomes paramount, especially in small-scale systems without classical simulation assistance. This pursuit unravels unique challenges in quantum and hybrid systems, crucial given quantum computing's inherent parallelism and nonlinear dynamics.\\

Quantum computers, although limited in computational capacity, are steadily advancing in the Noisy Intermediate-Scale Quantum~(NISQ) era~\cite{kim2023, robbiati2023real, arute2019quantum,joshi2023exploring, PhysRevX.13.041052, Barron2024}. They excel in certain areas, particularly quantum-classical hybrid algorithms~\cite{RevModPhys.94.015004}, which synergetically integrate quantum and classical sub-routines in the most efficient manner as is done in the Variational Quantum Algorithm~(VQA). In quantum machine learning, Quantum Neural Networks (QNN) compute cost functions using quantum hardware and optimize sub-routines on classical computers~\cite{Beer2020}. Empirical comparisons between QNN and classical neural networks (CNN) are essential due to the absence of analytical models for heuristic-based algorithms. Recognizing limitations, such as communication latency, quantum hardware precision, choice of optimizer, training duration and resource allocation, are crucial for these hybrid systems in the NISQ era.\\
In most of the current implementations of quantum machine learning algorithms, the training of the quantum machine has been performed either in a classical simulator~\cite{PhysRevLett.113.130503, PhysRevLett.114.110504} or with significant guidance from a classical machine~\cite{Havlíček2019}. In the later case, the quantum machine is trained only in the vicinity of an optimal solution. Currently, the small system size allows efficient classical simulation to guide the quantum machine in training, but eventually a large scale quantum-classical hybrid system needs to be trained by itself without any classical simulation guidance. Unlike classical machine learning, classical simulation guided supervised learning is unique to quantum systems. The challenges in implementing unguided training on a quantum machine are multi-fold: (a) transfer of classical data to a quantum computer, (b) drifts in operational parameters of a quantum computer, (c) the selection of an efficient classical optimizer, (d) latency inherent in the classical-quantum interface. In this letter, we present the implementation and comprehensive analysis of a training protocol for a quantum classifier, leveraging a hybrid architecture that combines an ion trap-based quantum system with a classical processor. Our implementation not only addresses the aforementioned challenges but also unveils additional considerations necessary for optimizing the efficiency and enhancing the performance of the hybrid system beyond that of classical solutions.\\

The choice of an appropriate classical optimizer is often associated with the specific nature of the problem and the topology of the hyper-parameter space. While this correlation is well-established, as evidenced by the phenomenon of barren plateau~\cite{McClean2018}, in the case of quantum systems in the NISQ era the challenges are compounded by the presence of noise and systematic biases. We show that a genetic algorithm~(GA), based on the concept of biological evolution, outperforms various gradient-based approaches when deployed on actual quantum hardware systems that are noisy. The GA considers the quantum hardware as a black-box with inherent noise and bias but as long as the drift of the operational point remains low, the training process converges rapidly. On the contrary, the measurement postulate of quantum mechanics forbids gradient measurement without affecting the intermediate states of the quantum computer. Consequently, gradient-based optimization strategies incur significant temporal and resource costs.\\
Uploading classical data to a quantum computer poses a major challenge, especially in machine learning applications where training dominates computational time. Previous Quantum Machine Learning (QML) implementations typically conduct training on a classical simulator, reserving the QML system for validation~\cite{Dutta-2022}. This is feasible due to the small size of NISQ quantum hardware, allowing classical simulation. Classical training data is usually mapped to quantum states' amplitude or phase~\cite{doi:10.1073/pnas.2006373117}, or it can in-principle be uploaded to a quantum random access memory~\cite{Park2019, PhysRevLett.100.160501}. While the former approach has been implemented on various platforms with specific circuit depth requirements~\cite{doi:10.1073/pnas.2006373117}, the later is yet to be experimentally realized due to technical complexity. In this study, we employ a third approach, directly uploading classical data into quantum gate parameters using the data re-uploading algorithm~\cite{perez2020data}. Previous work validated QNN using this algorithm~\cite{Dutta-2022}, although optimal machine learnable parameters (MLP) were derived from training on a classical simulator due to the cost and accuracy limitations of training on NISQ machines.\\%
In the following, we introduce a binary classification problem and show the result of training conducted on a hybrid quantum-classical~(HQC) system based on an ion-trap platform without the guidance of any classical simulation. The final accuracy and the rate of convergence of the three training routines are compared in a real experimental set-up. We argue that the GA outperforms gradient-based optimizers for this type of problem where the hyper-parameter space is replete with many local minima as seen in fig.~(\ref{fig:fig1}). It is also important to recognize that the performance of the classical sub-routine depends on the choice of ansatz used for the quantum sub-routine. To this end, we have conducted a comprehensive analysis of the hybrid system to ensure that the training process is both efficient and accurate. The section~(\ref{sec:meth},~\ref{sec:quantum-processor}) provides extensive details on the experimental procedures, highlighting key aspects of the classical and quantum hardware that are essential for the effective training of the hybrid system.\\
\section{Results}\label{sec:results}
\begin{figure}[ht]
% \begin{center}
\centering
\subfigure[\hspace{0mm} ]{
\includegraphics[width=0.2\textwidth]{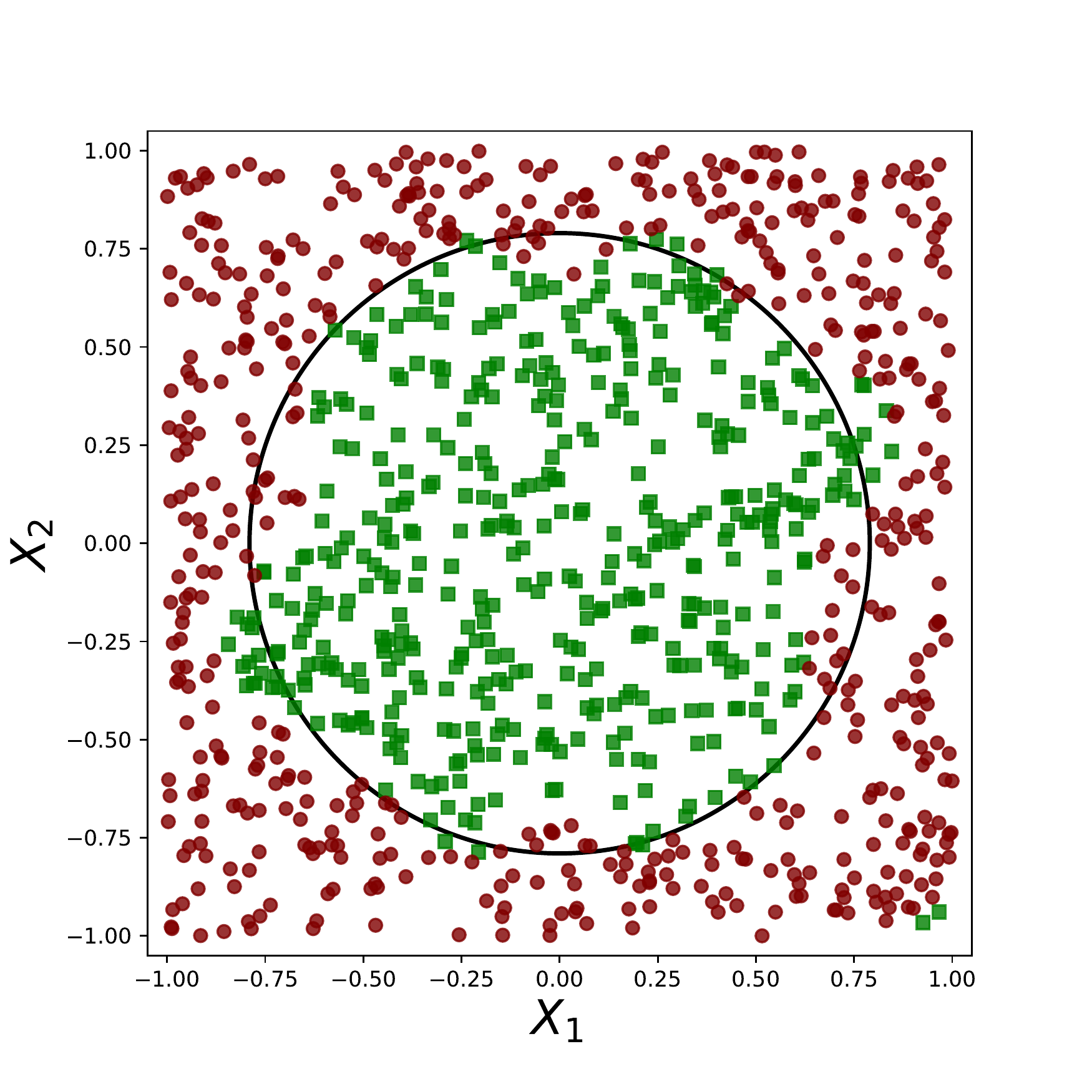}
 }
\vspace{-5 mm}
\subfigure[\hspace{0mm} ]{
\includegraphics[width=0.26\textwidth]{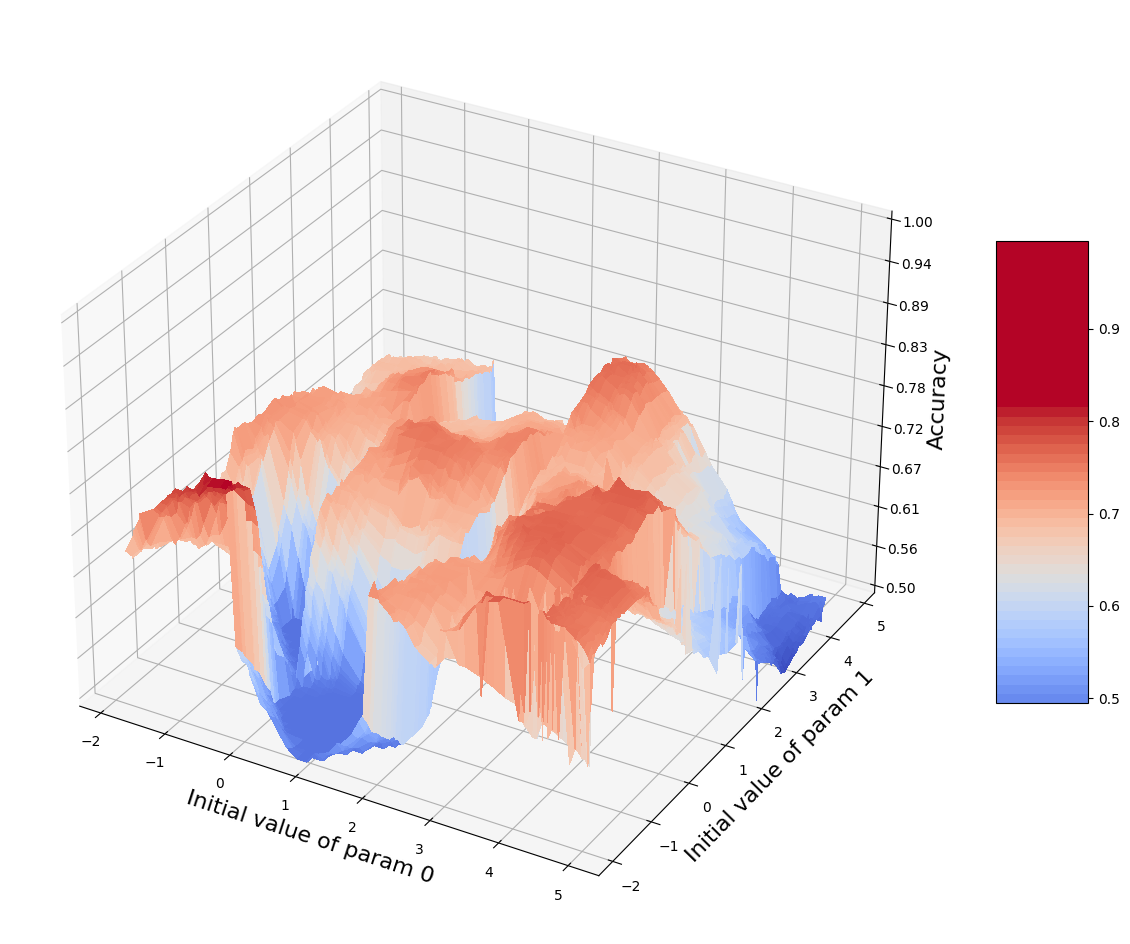}
 }
% \end{center}
\vspace*{-1mm}
\caption{(a) Binary classification problem statement: the task for the machine is to assign a label (brown/green) to any arbitrary point on a two dimensional space based on its coordinate whether inside or outside the pre-defined circle (black line) respectively. (b) Problem landscape: highest achievable accuracy in the neighborhood of a randomly selected parameter vector using Ansatz-2C, details of ansatz described in the section~(\ref{sec:kernel}). The optimal accuracy of the neighborhood is plotted against the first two parameter values. The landscape is filled with local minima and maxima, and tiny perturbation in one parameter could lead to sharp drop of the classifier's accuracy.}
% \caption{(a) Binary classification problem: assign a label (brown/green) to any arbitrary point on a two dimensional space based on its coordinate. (b) Problem landscape: highest achievable accuracy in the neighborhood of a randomly selected parameter vector using Ansatz-2C described in the supplement. The optimal accuracy of the neighborhood is plotted against the first two parameter values.
% }
\label{fig:fig1}
\end{figure} 

A classifier is an abstract map, $f : \mathbb{R}^n \to \mathbb{C} $, from the Euclidean space (where data resides) $\mathbb{R}^n$ to a finite set $\mathbb{C}$, which represents the individual classes we are trying to classify. The goal is to train the map $f$ by varying certain variables (also called machine learnable parameters) of the map, such that data of a certain type or label maps to region in $\mathbb{C}$ space which is clearly separable from data with all other labels.\\ 
In the data re-uploading quantum classifier paradigm, we define a sequence of unitary operations $U_l$ applied to an initial state $\ket{0}$, such that the final state $\ket{\phi}$ is:
\begin{equation}
\ket{\phi} = U_L(\bar{\theta}, \bar{x}) U_{L-1}(\bar{\theta}, \bar{x}) \dots U_0(\bar{\theta}, \bar{x}) \ket{0} = \prod_{l=0}^L U_l(\bar{\theta}, \bar{x})\ket{0} \
 \label{equation:abstract-classifier} 
\end{equation}

\begin{figure*}[ht]
\begin{center}
% \hspace{-9mm}
\subfigure[]{%
\label{fig:fig3a}
\centering
\resizebox{0.26\linewidth}{!}{
%%%%%%%%%%%%%%%%%%%
\begin{tikzpicture}[
        geom/.style={
            rounded corners,
            minimum width=25mm,
            minimum height=9mm,
        },
        state/.style={fill=blue!40!teal!40,geom},
        other/.style={fill=orange!40,geom},
        txt/.style={align=center,font={\small\ttfamily}},
        arrlbl/.style={blue,align=center}, 
        %
               % ~~~ new ~~~~~~~~~
        decoration={
            markings,
            mark=at position 0   with {\arrowreversed[black, scale=2]{Stealth}},% 
            mark=at position .15 with {\arrowreversed[black,scale=2]{Stealth}},
            mark=at position .35 with {\arrowreversed[black,scale=2]{Stealth}},
            mark=at position .5  with {\arrowreversed[black, scale=2]{Stealth}},% west
            mark=at position .75 with {\arrowreversed[black, scale=2]{Stealth}},% south
        }
]
        \draw[postaction={decorate}] (0,0) circle[radius=\rd];
        %
               % Blocks
        \node[other] (OP) at (  90:\rd) {\large Optimizer};
        \node[state, font=\large] (QP) at (  17:\rd) {$\rm R(\{\vec{\Theta}\},\{\vec{X}\})$};
        \node[state,minimum height=15mm] (QC) at ( -30:\rd) {
            \resizebox{.25\linewidth}{!}{
            \hspace{-.58cm}~\Qcircuit @C=.8em @R=.5em{
                \push{\cdots} &\gate{R_y} & \gate{R_z} & \gate{R_y} & \gate{R_z}&\cdots\\
                \protect\gategroup{1}{2}{1}{3}{0.3em}{--}
                \protect\gategroup{1}{4}{1}{5}{0.3em}{--}
            }
            }
    };
        \node[state] (QM) at ( 210:\rd)  {
        \resizebox{.16\linewidth}{!}{
    \hspace{.0cm}~\Qcircuit @C=.8em @R=.5em{
    \push{\hspace{1mm}} & \meter &\qw& \push{\cdot}
    &    
    }}
    };
    \node[other,font = {\large}] (PP) at ( 165:\rd) {$f(|\langle y | \phi \rangle |^2 )$ };
    \node[rounded corners, fill=blue!40!teal!40,font = {\large},minimum width=10mm] (QX) at ([shift=(0:2)]QP)       {$|\large 0 \rangle$};
    \node[rounded corners, fill=orange!40,font = {\large}] (DT) at ([shift=(60:2)]QP)     {Data: $\{\vec{X\}}$};
        % Connections
        \draw[->] (DT) -- (QP.25);
        \draw[->] (QX) -- (QP);

        % ~~~~~~~~~~~~~~~~~
        %
        \node[arrlbl, anchor=north west, black] at (PP.260) {\large Quantum state:\\$|\scalebox{1.3}{\large$\phi$} \rangle$};
        \node[txt, font= {\large}] at ([shift=(94:-1.)]QM) {\ \ Quantum\ \ \ \ state \\measurement};
        \draw[txt, decoration={text along path,reverse path,text align={align=center},text={|\large|Classical post-}},decorate] (.01,1.4)arc (0:260:1.55);
        \draw[decoration={text along path,reverse path,text align={align=center},text={|\large| processing}},decorate] (-.001,1.3)arc (0:260:1.1);%(.8,4.4)arc (120:170:8.5);
        \node[txt,font = {\large}] at ([shift=( -85:1.0)]QX) { Quantum\\pre-processing};
        \node[txt,font = {\large}] at ([shift=(100:-1.2)]QC) {\ \ \ Quantum circuit};
        \node[txt] at ([shift=(30:-1.2)]QC) {\rm Layer $\textit{i}$};
        \node[txt] at ([shift=(150:-1.2)]QC) {\rm Layer $\textit{i}+1$};
        \node[txt,font = {\large}] at ([shift=(138:1.5)]QP) {\\$\{\vec\Theta\}$};
        \node[txt,font = {\large},rotate=-49, sloped] at ([shift=(107:1.7)]QP) {Parameters:};
    \end{tikzpicture}
    }}
 % }%
\subfigure[\hspace{0mm}]{
\label{fig:fig3b}
\includegraphics[width=0.22\textwidth]{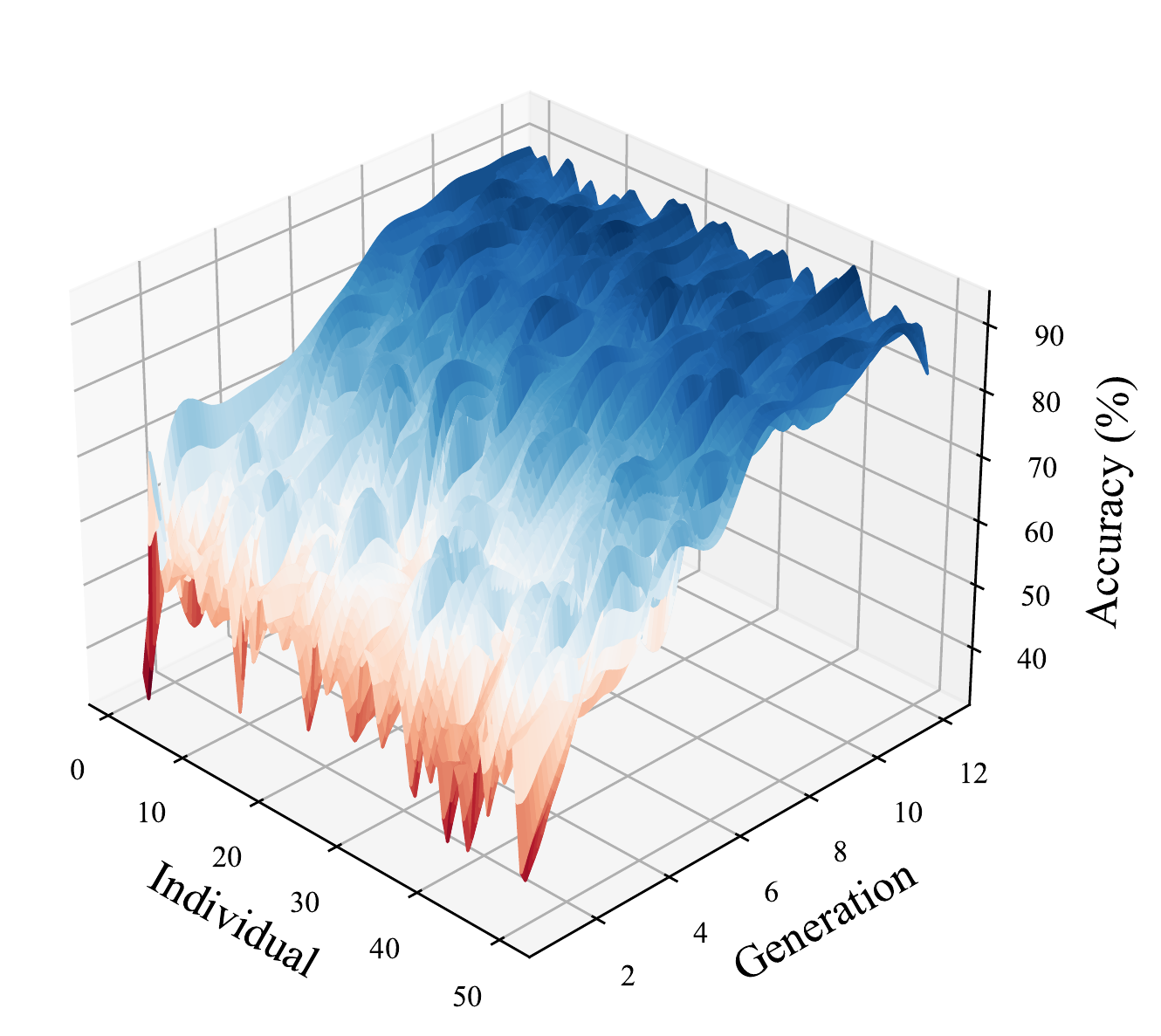}
 }
\subfigure[\hspace{0mm}]{
\label{fig:fig3c}
\includegraphics[width=0.22\textwidth]{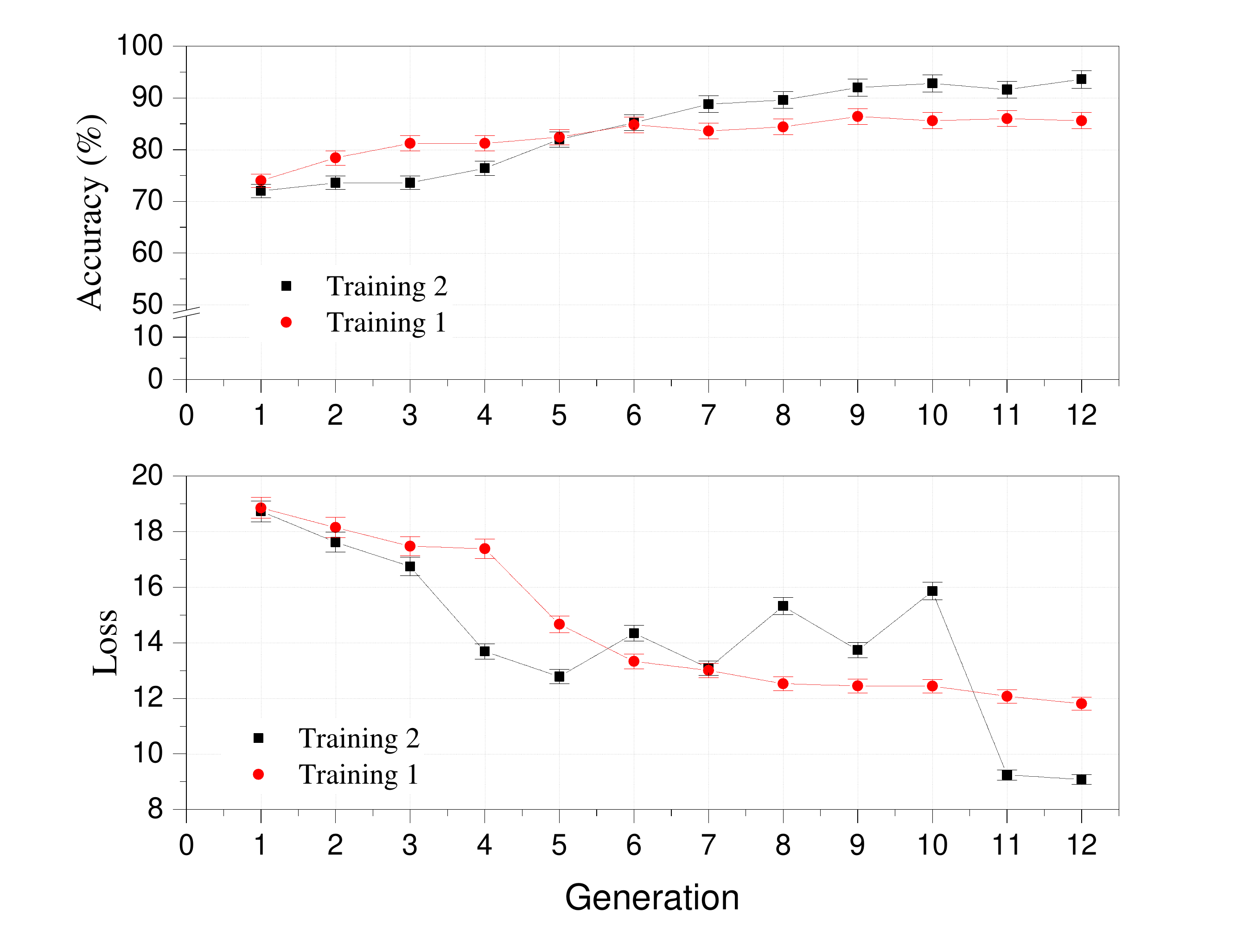}
 }
\subfigure[\hspace{0mm} ]{
\label{fig:fig3d}
\includegraphics[width=0.22\textwidth]{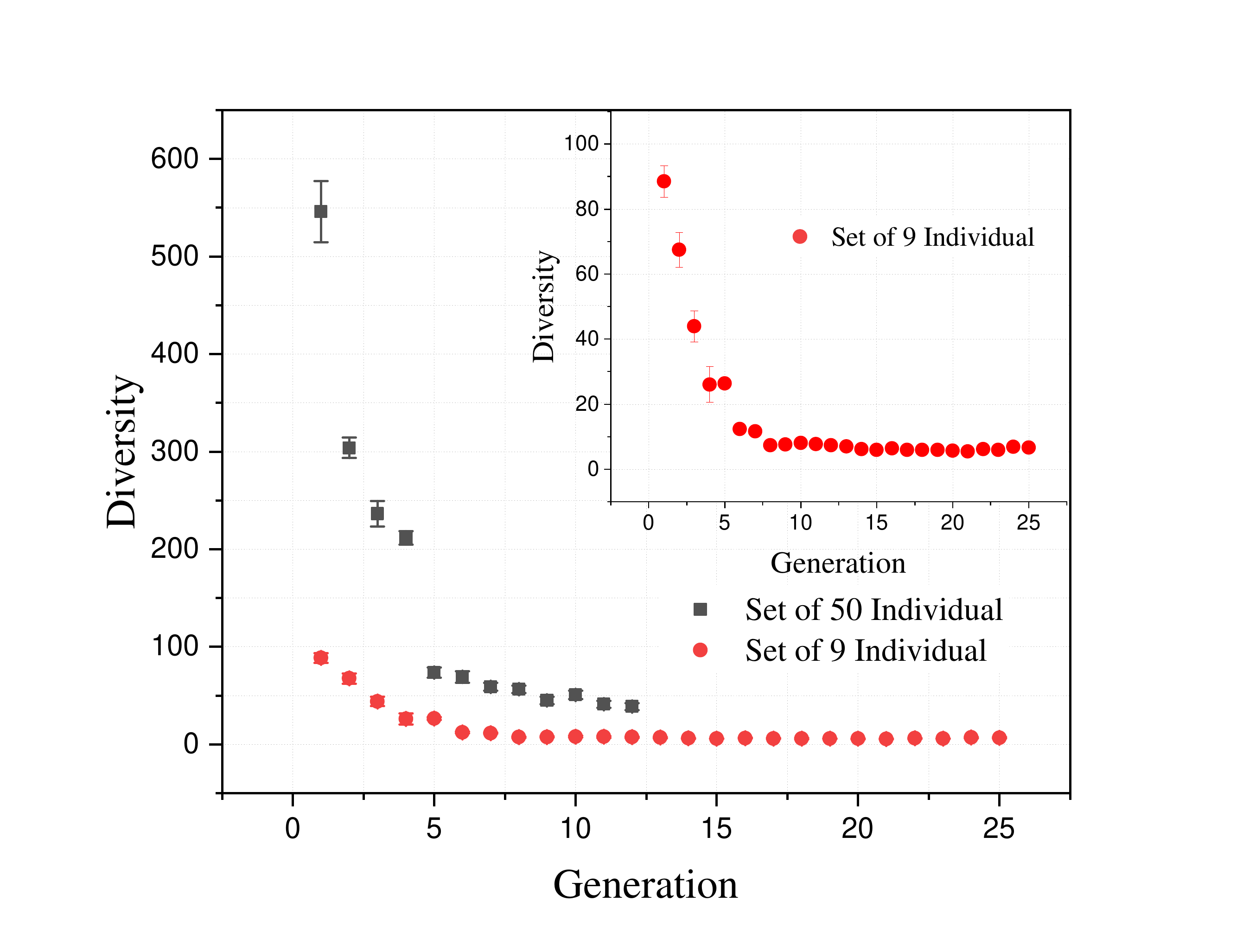}
 }
\end{center}
\vspace{-7mm}
\caption{ (a) This concise circular representation highlights the quantum training process for classification using classical and quantum components (boxes filled in orange and blue colour respectively). The flow illustrates the path of data and operations, starting with classical pre-processing, moving to quantum processing, and concluding with classical post-processing and model updating. Arrows indicate the direction of data flow and processing steps. See the text for details. (b) Search for optimal parameters for classification task using GA: The ion trap-based QPU is used for training on $250$ random data points. The depth of the circuit is kept fixed to $4$ layers as it is sufficient to classify the current problem with accuray above $90\%$. For this training, a set of $50$ individuals have been used and the data is uploaded using ansatz 2C. A detailed discussion is in the Appendix\ref{sec:kernel}. (c) The best accuracies and losses (cross entropy) for each generation from two independent experimental runs, labeled as Training 1 and Training 2, are displayed here. These experiments were performed using the same QPU. The best accuracy achieved is $93.6\pm 2\%$. (d) The diversity of individuals varies from one generation to the next. When starting with $50$ individuals, the initial diversity is significantly higher compared to beginning with only $9$ individuals. The inset figure provides a magnified view of the relationship between diversity and generation for a set of $9$ individuals. A higher initial diversity is associated with achieving higher accuracy with fewer generations. 
}
\label{fig:fig3}
\end{figure*}
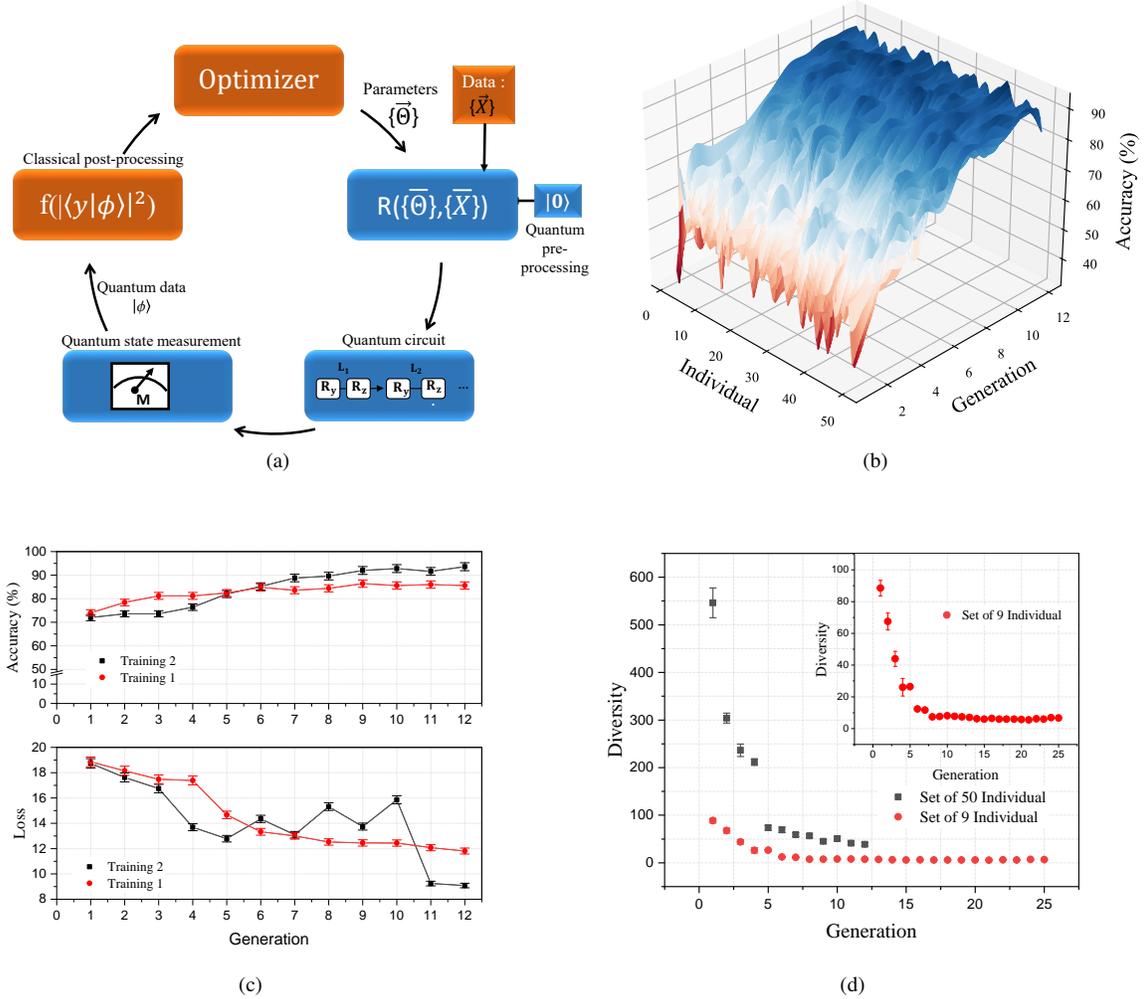

Our quantum circuit processes a vector $\bar{x} \in \mathbb{R}^n$, combined with variational parameters $\bar{\theta}$, creating a sequence of unitary operators applied to an initial state $\ket{0}$. The output state is measured and classified as $A$ (green squares in fig.~(\ref{fig:fig1})) or $B$ (brown dots in fig.~(\ref{fig:fig1})) based on predefined criteria in the Hilbert space. The classifier, utilizing the data re-uploading algorithm, has been benchmarked against CNNs~\cite{Dutta-2022}, showing promise despite limited parameter space availability. For training, we selected the circle as a test-bed boundary, offering a dense array of local minima, ideal for testing as depicted in fig.~\ref{fig:fig1}.\\ 
A set of $250$ random data (points in a 2D plane) is chosen for the training procedure. As illustrated in Fig.~\ref{fig:fig3a}, training involves a repetitive sequence of the following steps: (i) classical optimizer generating a set of variational parameters, (ii) mapping of data and variational parameters to gate parameters, (iii) sequential uploading of mapped parameters to the ion trap quantum computer, (iv) projection measurement on the quantum computer and cost function evaluation on the classical computer, (v) optimizing on classical computer and finally closing the loop by returning to step (i). The sequence (i-v) remains consistent throughout this work, with the only variation being in the classical optimizers. This approach allows for a direct comparison of each optimizer under the same conditions. One of the most crucial steps in this process is (ii) which differentiates the challenges in an analogue simulator~\cite{doi:10.1073/pnas.2006373117} to a circuit-based quantum computer like ours.

%%%%%%%%%%%%%%
%%%%%%%%%%%%%%%%%%%%%%%%%%%%
\begin{center}
\subsection{Genetic Algorithm}
\end{center}
The Genetic Algorithm (GA) is a family of optimization algorithms inspired by the process of natural selection, emphasizing the survival of the fittest principle. According to a survey~\cite{katoch2021review}, there are $4$ common degrees-of-freedom or hyper-parameters in a GA.
Each of these hyper-parameters as applicable to the quantum classifier has been investigated in appendix~\ref{sec:genetic-optimizer}. Once selected, these parameters remain unchanged throughout classifier training. The key training outcome is depicted in Fig.~\ref{fig:fig3c}. Using a population size of only $50$, a training accuracy exceeding $90\%$ is achieved just within nine generations of the training routine. The figure presents two experimental sample training runs, illustrating the evolutionary process and its robustness. The evolution of the accuracy with generations denoted by the $\protect\tikzbullet{red}{red}$ points initially improves until the $6^\text{th}$ generation but then remains at $83\%$ even after the $10^\text{th}$ generations, while $\blacksquare$ points reach over $90\%$ accuracy following a different path. The uncertainty of experimental data is $\pm 2\%$. The error represents the standard deviation of $10$ repeated trials conducted on the same dataset, indicating that underlying systematic uncertainties contribute to the overall uncertainty in the accuracy, as detailed in~\cite{Dutta-2022}. Additionally, the plot illustrates the evolution of cross-entropy loss across generations, indicating iterative refinement and improved accuracy alignment.

The experimental training run (shown in $\blacksquare$ in Fig.~\ref{fig:fig3c}) is further analyzed in terms of the individuals' (in a population of 50) evolutionary path over generations, as illustrated in Fig.~\ref{fig:fig3b}. The colormap shows the training accuracy of each individual in the population for a given generation, starting from a wide range of distribution (denoted by the spread of the colour). To facilitate an easier interpretation of the colormap, the values between discrete measurement points (at integer values of population number and generation number) are interpolated and smoothed. Note that the GA in each generation creates a new set of population based on the algorithm and hyper-parameter values, as discussed earlier.\\
When the mutation probability is held constant, the optimizer tends to explore only local optima, resulting in a lower final accuracy compared to scenarios where the mutation probability is dynamically adjusted based on the generation number (fig.~\ref{fig:analysis-of-ga-hyperparams}). This phenomenon can be attributed to the inherent nature of GAs, where a diverse population allows for a more global search across the solution landscape. Conversely, maintaining a static mutation rate throughout the learning process impedes the algorithmic convergence, leading to sub-optimal exploration (sec.~\ref{sec:genetic-optimizer}). In the context of the quantum-classical hybrid system, the large exploration space as well as the dynamical adjustment of the hyper-parameter helps in dynamically mitigating any bias that arises in the quantum system over time.

Contrasting with the accuracy attained by GA runs with a population of $50$, the maximum accuracy achieved by runs with a smaller population of just $9$ was approximately $85\%$, even after the $25^\text{th}$ generation. The landscape depicting the evolution of accuracies over generations for this individual is presented in fig.~\ref{fig:ga-9-pop} in the section~(\ref{sec:genetic-optimizer}).
\\
Having a lower number of individuals as shown in fig.~\ref{fig:fig3d} introduces a different set of challenges. In our experiment, we observed that the rate of convergence differs when comparing populations of $9$ and $50$ individuals per generation during training. With a population of $9$, diversity is initially low, which leads to rapid convergence; however, both the accuracy and final diversity are inferior to what is achieved with a population of $50$. This suggests that smaller populations may lead to stronger exploitation of the search space at the expense of exploration.

So far, we have been discussing the impact of GA on the rate of convergence of a hybrid quantum-classical training system. However, the most commonly used (classical) optimizer for these systems is based on gradient descent~(GD) algorithm. In the following, we show the learning performance of the HQC using a few flavors of GDs. 
\\
%%%%%%%%%%%%%%%%%%%%%%%%%%%%

\subsection{Gradient-based Algorithms}

Gradient-based method is an iterative optimization method that leverages on first and possibly higher order derivatives of the cost function with respect to machine learnable parameters. Here, we explore the application a quasi-Newton method, specifically the Broyden-Fletcher-Goldfarb-Shanno (BFGS) algorithm~\cite{10.1093/comjnl/7.2.149}, which is being utilized for the first time in a hybrid quantum-classical system related to a quantum classifier. We focused our studies to the BFGS algorithm and tested it with two different gradient evaluation methods as discussed in the following sections. Gradient Descent and Stochastic Gradient Descent (SGD) are well-established techniques in the fields of statistics and classical deep learning and have been previously employed in optimizing random quantum circuits~\cite{PhysRevA.98.032309}. However, within our NISQ setup, these methods have demonstrated less reliable convergence and hence only discussed in the section~(\ref{gradient}) as a comparison.

The BFGS method is a prominent iterative technique for tackling unconstrained nonlinear optimization problems. It incorporates second-order derivative information to guide the optimization path, resulting in a more reliable convergence trajectory compared to methods that rely solely on first-order gradients. As a member of the quasi-Newton family, BFGS approximates the Hessian matrix, avoiding the computational expense of calculating the Hessian directly. This cost-effectiveness is especially beneficial for training routines on our NISQ systems. In contrast to classical machine learning algorithms, where gradients are typically computed analytically via back-propagation, quantum systems necessitate measuring the quantum processor to obtain gradients with respect to specific parameters. In the following sections, we discuss the implementation of two distinct methods of gradient estimation on a QPU within a $14$-dimensional parameter space designed for machine learning. We then conducted a comparative analysis of the performance of these two gradient estimation techniques.

%%%%%%%%%%%%%%%%%%%%%%%%%%
\begin{figure}[ht]
% \hspace{-2 cm}
% \begin{center}
\includegraphics[width=0.45\textwidth]{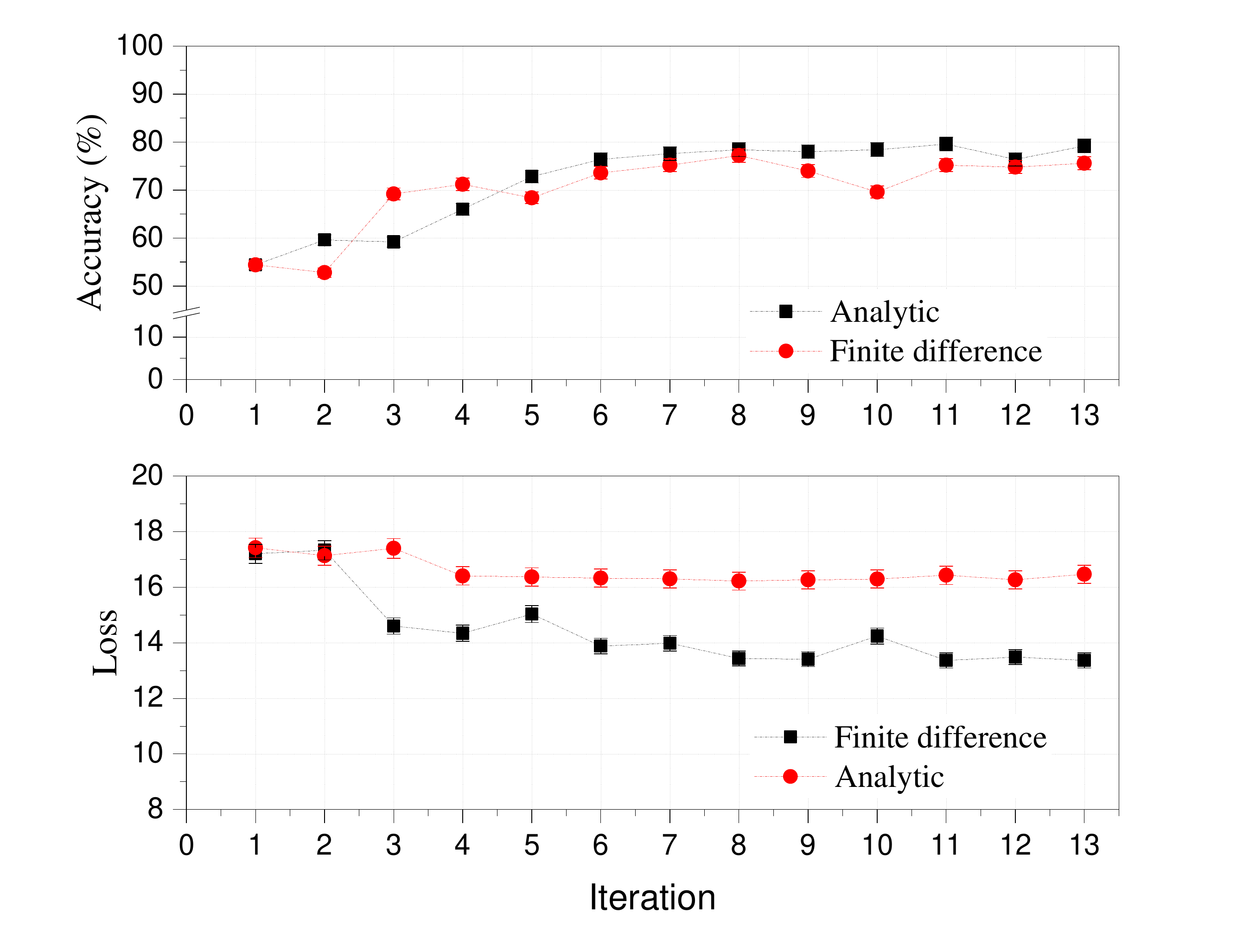}
% \end{center}
\caption{The training accuracy (top) and cross-entropy loss (bottom) on the QPU are graphed against iteration steps, utilizing the BFGS optimizer for both finite difference and analytic approaches. The search for the best parameters for the classification task employed the BFGS-finite difference method with a step size of $0.5$. Connected lines between data points are utilized in tracking the data's progression only. Here, we have presented data up to 13 iterations for both the finite difference and analytic methods to facilitate comparison. In reality, we have collected data up to approximately 30 iterations (~600 Measurements) to corroborate the measurement with GA, as shown in Plot~\ref{fig:compare-optimizers}}
\label{fig:fig5}
\end{figure} 
% %%%%%%%%%%%%%%%%%%%%%%%%%%%%%%%%%
% % %
\begin{figure*}[t]
\begin{center}
\subfigure[\hspace{-1 mm}]{
\label{fig:compare-optimizers}
\includegraphics[width=0.42\textwidth]{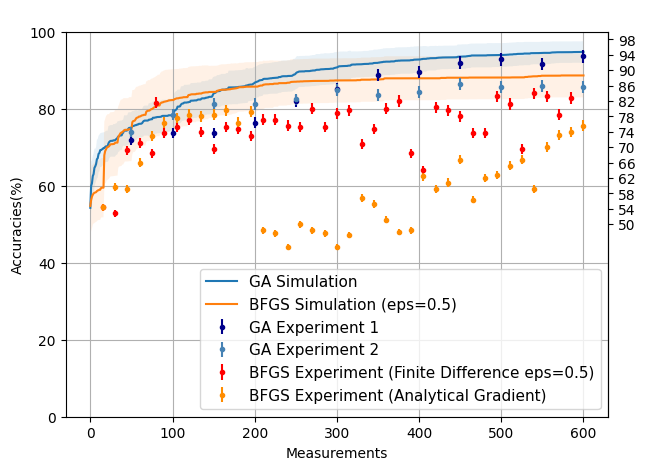}
 }
% \vspace{-5 mm}
\subfigure[\hspace{0 mm} ]{%
\label{fig:compare-optimizers-simulation}
\includegraphics[width=0.42\textwidth]{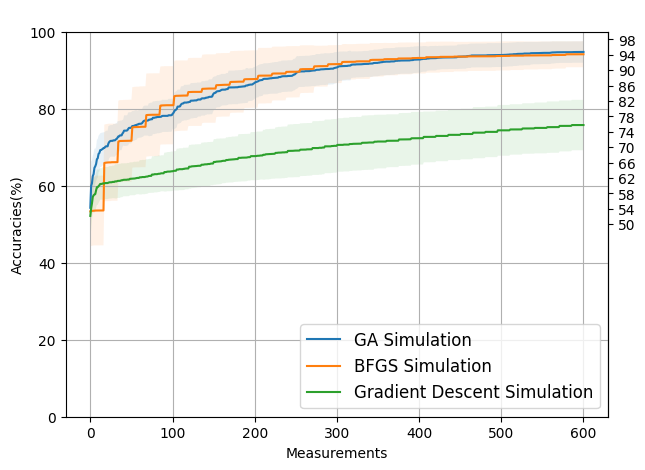}
 }
\end{center}
\vspace{-5 mm}
\caption{ (a) Comparison between simulation and experimental results for Genetic Algorithm and BFGS: The BFGS simulation run was modified to use a larger step size of 0.5 in the finite difference gradient approximation step in order to match our experimental setup. BFGS-based experimental runs suffer from precision issues due to system noise as explained in the sec.~\ref{gradient}. Referring to the problem landscape illustrated in fig.~\ref{fig:fig1}, we can see that small perturbation from noise could significantly alter the direction and magnitude of our gradient. GA-based runs are more insensitive to noise and converge similarly as is in the case of simulation.%
(b) Simulation comparison between genetic and gradient-based optimizers: Gradient descent using first order derivative seems to converge less rapidly in this setup. L-BFGS-B variant was used and it gives comparable performance to Genetic Algorithm. Plots for simulation consist of the average (solid lines) and one standard deviation (semi-transparent band) across multiple runs. Plots for experiment consist of individual runs (solid dots) and experimental uncertainties (error bars).}
\end{figure*}
% %%%%%%%%%%
\subsubsection*{Methods of Gradient Estimation}

The BFGS method is renowned for its superior convergence properties. In order to leverage this algorithm, it remains to estimate the gradient at each desired $\bar{\theta}$ (see section~\ref{gradient}) value. In classical machine learning problems, BFGS-based methods usually approximate the derivatives using the finite difference approach.\\

Generally, any gradient-based algorithms will perform better when the step size of finite difference computation is small. In our problem setting, we found through simulation that a step size below $0.05$ is needed. However, achieving such a small step size is challenging in practice with the NISQ devices. To address this issue, we employ a different technique known as the \textit{parameter shift method}~\cite{crooks2019gradients}. This technique evaluates gradient of any parameter using arbitrary shifts in two opposite directions. We fix shift sizes of $\pm\frac{\pi}{2}$ in order to maximally leverage this technique and reduce the effect of random noise. An alternative method, particularly applicable in our context of data re-uploading algorithm is \textit{analytic gradient}. It is possible to find the analytic gradient of our circuit at each point in the parameter space, thus using the QPU, the analytic gradient can be measured without any approximation and iterated successively.

Figure~\ref{fig:fig5} illustrates  the training accuracy and cross-entropy loss of our binary classifier, comparing the finite-difference and analytic gradient methods, distinct from the genetic algorithm. The finite difference gradient method yields lower loss values compared to the analytic gradient approach. By estimating gradients through differences of measurements at paired points, it partially cancels correlated noise and bias. In contrast, the analytic gradient method, relying on single measurements per term, lacks this noise-cancellation effect, making it more sensitive to hardware noise. This underscores the influence of noise characteristics on optimization performance and the need to consider hardware-specific noise when designing optimization methods for QML applications. Notably, after the $6^\text{th}$ iteration, both gradient based methods exhibit stagnant accuracy improvements, mirrored in the  cross-entropy loss, where its slope remains consistent over subsequent iterations. In contrast, the GA demonstrates a pronounced decline in slope across generations as in fig.~(\ref{fig:fig3c}). 
We compare CPU-based simulation results with QPU-based experimental outcomes as shown in Fig.~\ref{fig:compare-optimizers}, applying the same operational conditions to finite-difference, analytic gradient measurement techniques and GA. Notably, the simulations represent an idealized scenario devoid of noise or bias, which is not the case for the experimental runs. Consequently, the observed discrepancies in the learning curves between simulation and experimental results can be attributed to both bias and noise inherent in the quantum system. In our earlier work~\cite{Dutta-2022}, we demonstrated that employing a classically bootstrapped learning method on the QPU can enhance the performance.

Overall, the discrepancies observed between the simulation and experimental results cannot be solely attributed to experimental bias. Given that the BFGS method is deterministic, unlike the probabilistic nature of GA, a closer alignment between the datasets is anticipated. However, the experimental data points consistently exhibits lower accuracy compared to the idealized, noise-free simulation. This divergence is likely due to factors such as bias, decoherence and inherent noise within the quantum subroutine. To mitigate these issues, we have explored post-measurement error correction techniques, which are detailed in the section~(\ref{sec:residual-error}). However, the GA is robust, in particular with respect to bias which is evident from the good match between simulated trajectory (blue defused band) and the two experimental runs ({\color{blue}$\bullet$}). On the contrary, we observe that the BFGS gradient experiments ({\color{orange}$\bullet$}, {\color{red}$\bullet$}) under-performs as compared to the CPU simulation (orange defused band). One interesting observation in the BFGS analytic gradient experiment is the sudden drop in the accuracy around $200^\text{th}$ measurement which we attribute to the existence of the cliffs in the parameter space as shown and discussed in fig.~(\ref{fig:fig1}) at which a small bias in the experiment leads to a sharp fall in the gradient, followed by a gradual recovery. In some conditions we have also observed this in the ideal simulations. Finally, as a comparison of ideal simulation with smallest step-size ($0.001$) we show in fig.~(\ref{fig:compare-optimizers-simulation}) that both GA and BFGS are comparable in terms of achievable accuracy and the time to converge but gradient descent shows slow convergence. This indicates that BFGS methods are only as good as GA if and only if the QPU is noiseless.

Our findings indicate that the final training accuracy is largely unaffected by the choice of gradient evaluation method, whether analytical or finite-difference-based. In both instances, the final accuracy falls short of that achieved by GAs. It is also worth noting that the data re-uploading algorithm, which employs rotational gates dependent on sine and cosine functions, allows for analytic gradient evaluation that is less resource-intensive compared to other variational quantum algorithms.

After implementing both GA and gradient-based optimizers for HQC in the NISQ era, a cross-algorithm performance comparison is warranted. Key performance metrics for classifier training should include the rate of convergence, the final achievable accuracy, and the associated resource requirements. We will discuss each of these performance indicators in the following section.
%%%%%%%%%%
\begin{figure}[ht]
    \centering
    \subfigure[\hspace{0mm} ]{
        \includegraphics[width=0.45\textwidth]{ga_varying_layer.png}
        \label{fig:ga_varying_layer}
        }
    \vspace{0 mm}
    \subfigure[\hspace{0mm} ]{
        \includegraphics[width=0.45\textwidth]{bfgs_0.5_varying_layer.png}
        \label{fig:bfgs_0.5_varying_layer}
        }
    \caption{Comparative study of a binary classification problem using noiseless and noisy ($\sim \mathcal{N}(0,0.05)$, matching our experimental setup) quantum computer: (a) Genetic Algorithm solver with a population size of $50$, scattered crossover, exponentially decaying mutation rate, and SSS scheme. Details of the population size, mutation rate, and SSS scheme are provided in Appendix Sec.~\ref{sec:genetic-optimizer}. (b) L-BFGS-B solver with a step size of $0.5$ (for gradient estimation) used for the experiment. The experimental results are shown in Figure~\ref{fig:compare-optimizers} for comparison.
    }
    \label{fig:fig-extra}
\end{figure}
%%%%%%%%%%%%%%%%%%%

\section{Discussion}
\label{sec:sec4}

In classical machine learning frameworks, the convergence rate of the machine learnable parameters is characterized by the rate at which the algorithm refines its parameter estimates, as determined by the cost function. In contrast, the comparison of optimization methods such as GA and gradient-based optimizers in HQCs necessitates a precise definition of a ``computational cycle." For the purposes of this study, we define a computational cycle in the context of HQC as the rate of model improvement per projection measurement, where each measurement consists of 100 experimental realizations, commonly referred to as ``shots".

Empirical results from our comparative analysis clearly demonstrate that GA optimizers surpass gradient-based optimizers in terms of both convergence rate and final accuracy under our problem setup using NISQ devices. 
To further illustrate these differences, we include additional numerical experiments. In the appendix fig~\ref{fig:fig-extra1}, we compare the performance of a typical GA solver with that of the L-BFGS-B solver, evaluated under various step sizes for finite-difference gradient estimation.  While L-BFGS-B solvers perform well in noise-free environments with smaller step sizes (Fig.~\ref{fig:bfgs_0.005_varying_noise} and Fig.~\ref{fig:bfgs_0.05_varying_noise}), their performance drastically degrades with the introduction of noise. This is expected as the absolute magnitudes of the gradients have magnitudes similar to those of the noise level. 
% Fig~\ref{fig:fig-extra} showing the superirority of GA over gradient estimation where we consider the performance of the GA solver and the l-BFGS-b solver under Gaussian noise of $0.05$, matching our experimental setup, and place the noise-free results alongside the noisy simulation results. We can see that there's a visibly larger gap between noise-free and noisy results for the l-BFGS-b solver. }
%
Fig.~\ref{fig:fig-extra} further highlights the superiority of GA over gradient-based optimization by comparing the performance of the GA solver and the L-BFGS-B solver under Gaussian noise of $0.05$, consistent with our experimental setup. We also include noise-free results alongside the noisy simulations, revealing a visibly larger performance gap for the L-BFGS-B solver in the presence of noise.
\\
A critical factor influencing the performance limitations of gradient-based methods is the resolution of the step size. In the context of the binary classifier, the MLP parameter space is replete with local minima, which poses a significant challenge for gradient-based optimization. In contrast, GAs leverage a population-based approach to explore multiple regions of the MLP space concurrently. This diversity allows for a more comprehensive search and exploitation of multiple promising regions. Consequently, GAs exhibit a degree of robustness against the noise and systematic biases that may be present in the quantum subroutine, which are factors that can detrimentally affect the performance of gradient-based methods. The phenomenon of \textit{Barren Plateaus} presents a significant challenge in the field of variational quantum computing, where the optimization landscape becomes exponentially flat, hindering the training of quantum circuits. This issue is often mitigated by narrowing the search space of the problem \cite{cerezo2023does}. We wish to highlight that the intrinsic capability of GA to concurrently explore multiple regions of the search space renders them less susceptible to the Barren Plateaus problem \cite{acampora2022training}.

We now focus on the error mitigation strategy utilized in our implementation. As depicted in fig.~(\ref{fig:training-dataset}), a correlation plot contrasts the simulated data with the experimental outputs of the cost function across each training iteration. Under ideal conditions, this relationship would be characterized by a linear correlation with a slope of $45^\circ$ (represented by a dashed line). However, our observations indicate a dispersion of data points around this idealized line, accompanied by an apparent offset and altered slope. A histogram of the deviations from the ideal linear relationship reveals that the experimental error is predominantly Gaussian noise. Further analysis suggests that the altered slope of the correlation curve may be attributable to qubit decoherence, which introduces an undesirable systematic bias. To rectify this bias, we propose the utilization of the inverse correlation matrix as a compensatory measure applied to the experimental data. Despite this corrective approach, we note that it does not influence the gradient estimations in the context of binary classification problems, which follows from the mathematical definition of our cost functions as listed in Equations~(\ref{eq-cf1}-~\ref{eq-cf3}) in the section~(\ref{sec:cost-functions}). A thorough study of this error mitigation technique and its implications is provided in the section~(\ref{sec:residual-error}). Furthermore, we would like to point out that the ideal step size required to improve on the gradient-based approaches can be derived from the algorithmic error, setting the limits on the resolution of the parameter step size that can be practically applied to find the gradient as illustrated in fig.~(\ref{fig:gradient-analysis}).

In a HQC system, establishing quantum advantage fundamentally hinges on demonstrating superior speed or energy efficiency. However, quantifying these attributes presents several challenges: (a) The heuristic nature of neural network-based machine learning models precludes a straightforward assessment of computational resources from complexity theory; (b) In a hybrid system, the total computation time must account for the communication latency between the quantum and classical sub-routines, which can significantly impact overall performance; (c) The assessment of energy efficiency must encompass both quantum and classical sub-routines. However, the lack of a standardized framework for comparing these disparate computational architectures complicates this task, as they utilize fundamentally different physical resources. Despite these challenges, our research aims to demonstrate the capabilities of an ion trap-based HQC system in conducting supervised training with a data re-uploading paradigm, achieving classification accuracy exceeding 90\%. Presently, the computational time within our system is predominantly constrained by the communication overhead between the quantum and classical sub-routines. Potential improvements could be realized by integrating more substantial classical memory and computational resources proximal to the quantum hardware controls. Nonetheless, such enhancements may not be universally applicable, particularly in quantum computing architectures that operate at cryogenic temperatures.

In summary, we conclude that an ion trap-based hybrid quantum-classical NISQ processor is trainable by successfully executing supervised learning tasks for classification, without guided by classical simulation, achieving final accuracy exceeding 90\%. Moreover, through comparative analysis, we have identified the genetic algorithm as the superior classical optimizer subroutine for the data re-uploading classifier in the context of current NISQ processors, attributed to the limited precision in parameter settings inherent to NISQ technology. Lastly, we emphasize the significance to consider communication overhead between quantum and classical components in the overall performance evaluation, as it presents a potential bottleneck in hybrid NISQ processors.
% %%%%%%%%%%%%%%
%%%%%%%%%%%%%
\bibliography{main}
%%%%%%%%%%%%%%%%%%%%%%
\appendix
\newpage
% %%%%%%%%%%%%%%
\section*{Appendix}
\tableofcontents
%%%%%%%%%% Prefix a "S" to all equations, figures, tables and reset the counter %%%%%%%%%%
\setcounter{equation}{0}
\setcounter{figure}{0}
\setcounter{page}{1}
% \makeatletter
\renewcommand{\theequation}{A\arabic{equation}}
\renewcommand{\thefigure}{A\arabic{figure}}
\renewcommand{\thesection}{A\arabic{section}}
\renewcommand{\citenumfont}[1]{A#1}
% %%%%%%%%%% Prefix a "S" to all equations, figures, tables and reset the counter %%%%%%%%%%
\vspace{10pt}

Here, we provide a detailed description of the quantum classifier based on the data re-uploading algorithm. To facilitate a clear understanding of the performance of a quantum-classical hybrid algorithm in training a quantum classifier, binary classification of points on a plane is considered here for a circular boundary. The approach is based on data re-uploading algorithm~\cite{perez2020data} which is a quantum equivalent of a classical neural network (CNN). Like any machine learning protocol, the classifier has two modes of operation namely, training and validation. The training is performed on a random set of labeled data to obtain the variational parameters, alternatively called the machine learnable parameters~(MLP) of the classifier. Once successfully implemented, these MLPs are used to operate the classifier such that it is able to correctly label any random data into the right class. This mode of operation is called the validation.\\

Following the $\textit{Universal Quantum Circuit Approximation}$, we may approximate any classification function up to arbitrary precision by using sufficiently many unitary operations and with the data re-uploaded to each operation. Due to practical reasons, it is more convenient to represent arbitrary unitary operation as a combination of $R_y$ and $R_z$ non-commuting rotational gates of a single qubit. We also note that there are infinitely many ways of combining parameters ($\bar{\theta}$) and data ($\bar{x}$) (think the set of all polynomials with $2$ variables for example). For practical purposes choosing $\bar{\theta}\in \mathbb{R}^4$ and linearly combining them with $\bar{x}$ worked sufficiently well. It acts as kernel or feature space mapping for the quantum classifier. The quantum classifier circuit used for this comparative study is formed of $4$ layers. Thus from eq. (1),

\begin{align}
    \ket{\phi} = U_3(\bar{\theta}, \bar{x}) U_2(\bar{\theta}, \bar{x}) U_1(\bar{\theta}, \bar{x}) U_0(\bar{\theta}, \bar{x}) \ket{0},\\
    \ket{\phi} = \prod_{l=0}^4 U_l(\bar{\theta}, \bar{x})\ket{0}.\\
\end{align}

and each $U_l$ consists of an $R_y$ rotation followed by an $R_z$ rotation, whose values are computed according to the particular choice of the kernel. A good choice of the kernel is discussed in sec.~(\ref{sec:kernel}). Given a kernel with $L$ layers and parameters $\bar{\theta}$, a data point $\bar{x}$ and a label $y$, the measurement function returns the expected value projected onto the label state $\ket{y}$:

\begin{align}
    M(\bar{\theta}, \bar{x}, y) = \bigl| \bra{y} \prod_{l=0}^L U_l(\bar{\theta}, \bar{x})\ket{0} \bigr|^2.
\end{align}

In the following, we will briefly discuss the quantum classifier training methodology, the experimental set-up followed by a detailed comparative study of the importance of choice of anzats, optimizers and hyper-parameters.  

\section{Quantum classifier training methodology}
\label{sec:meth}

Similar to any machine learning protocol, the quantum classifier has two modes of operations namely, training and validation. The supervised training is performed on a random set of labeled data to obtain the variational parameters of the classifier. Once successfully implemented, these optimal parameters are used to operate the classifier such that it is able to correctly label any data into the right class. This mode of operation is called the validation.\\

Here, we operate the hybrid system in the training mode to obtain the optimal set of variational parameters. The quantum processing unit is based on ion-trap qubit, encoded in the electronic states of barium ion. In the following, We demonstrate the sequence carried out to optimize a predefined cost function by varying the values of the quantum gate parameters.\\

%%%%%%%%%%%%%%%%%%%%%%%%%%%
\begin{figure}[]
\begin{center}
\includegraphics[width=0.5\textwidth]{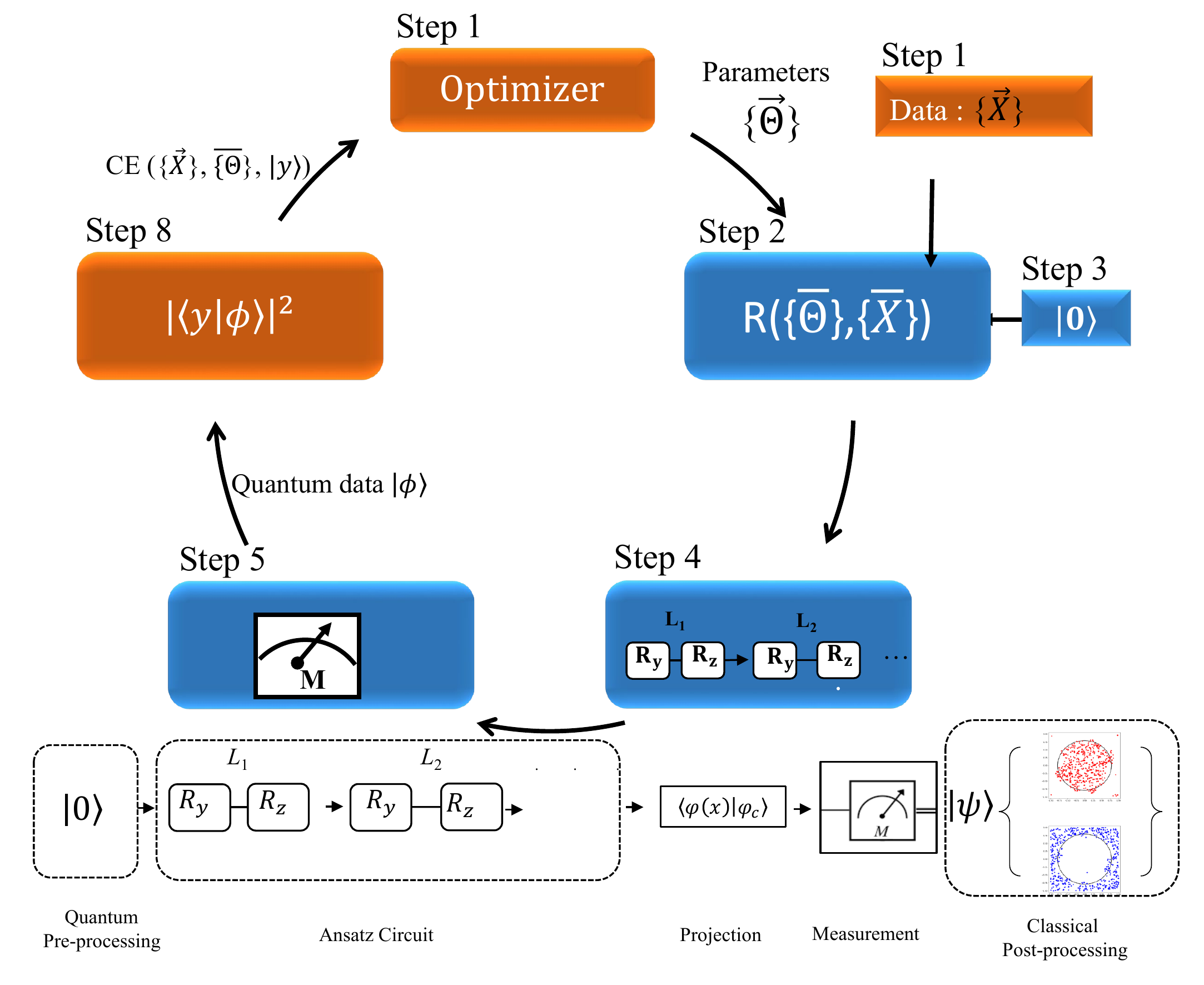}
\end{center}
\caption{Training: this concise circular representation highlights the quantum training process for classification using classical and quantum components. The circular flow illustrates the path of data and operations, starting with classical pre-processing, moving to quantum processing, and concluding with classical post-processing and model updates. Arrows indicate the direction of data flow and processing steps. The key points in this schematic are following: \textit{Classical pre-processing:} The CPU pre-processes the training data, and the resulting features are used as input to the quantum circuit.\textit{Classical to quantum interface (Encode)}: The classical data is encoded into a quantum gate using DDS and FPGA components (\textit{Quantum hardware}) to implement precise quantum control signals and quantum gates.\textit{Quantum circuit}: The QPU executes a quantum circuit to perform the classification task using the quantum-encoded data. \textit{Quantum to classical interface (Decode)}:  The quantum results are decoded back into classical data using PMT and FPGA. \textit {Classical post-processing}: The CPU processes the classical results, performs analysis, and updates the model parameters based on the quantum data.
}
\label{fig:training}
\end{figure} 
%%%%%%%%%%%%%%

\begin{itemize}
    \item[Step 1:] A set of initial randomize parameters of the circuit are generated in a classical computer. These parameters are convoluted with $250$ labeled training data following different possible \textit{Ansatz} (see sec.~(\ref{sec:kernel})) to transform the data to higher dimensional parameter space. This kernel data in the data re-uploading algorithm is uploaded to the quantum computer as gate parameters.
    \item[Step 2:] The list of gate parameters are loaded to the quantum processor. 
    \item[Step 3:] The state of the quantum processor is initialized to the ground state.
    \item[Step 4:] The list of gate parameters are sequentially applied as per the circuit. 
    \item[step 5:] Final state after execution of the circuit is projected on the label state and measurement of the projection is performed.
    \item[Step 6:] Steps 3-5 are repeated $150$ times to obtain the expectation value of the projection for one out of $250$ labeled training data.
    \item[Step 7:] Steps 2-6 are repeated for $250$ labeled training data points.
    \item[Step 8:] A suitable cost function is evaluated from the results in Step 7 which is fed back to the classical optimizer to provide the next set of parameters of the circuit.
    \item[Step 9:] Once the cost function reaches a pre-defined threshold, the algorithm stops and the final set of parameters is considered as optimal set.
\end{itemize}
The steps $2-6$ are carried out with the trapped ion quantum processor while the rest are performed on a classical computer. Since the algorithm runs in a loop between the quantum and classical processors, the communication time plays an important role. In the following, we will discuss the key steps and their implementation methods.
%%%%%%%%%%%%%%
\begin{figure}[ht!]
\begin{center}
\includegraphics[width=.45\linewidth]{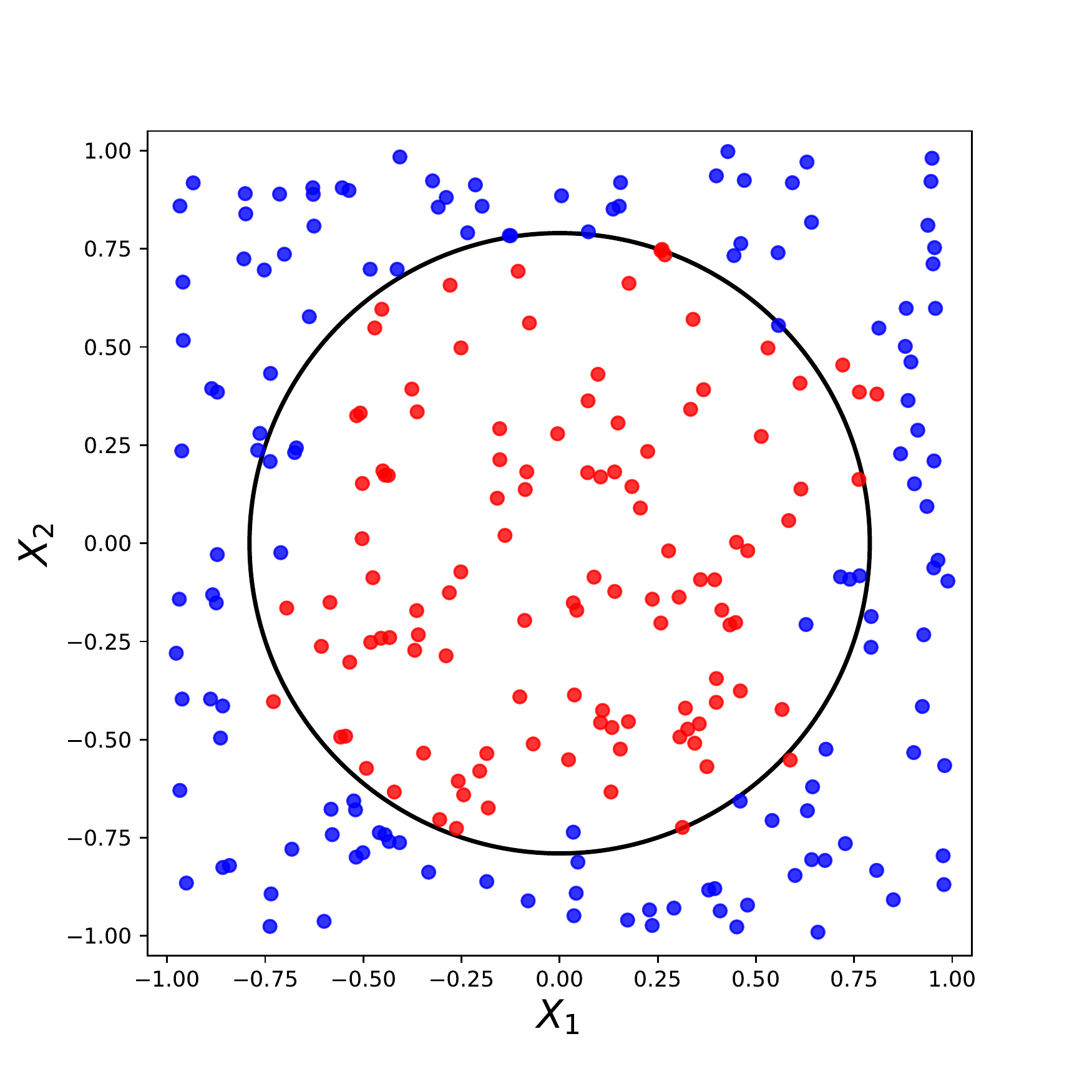}
\includegraphics[width=.45\linewidth]{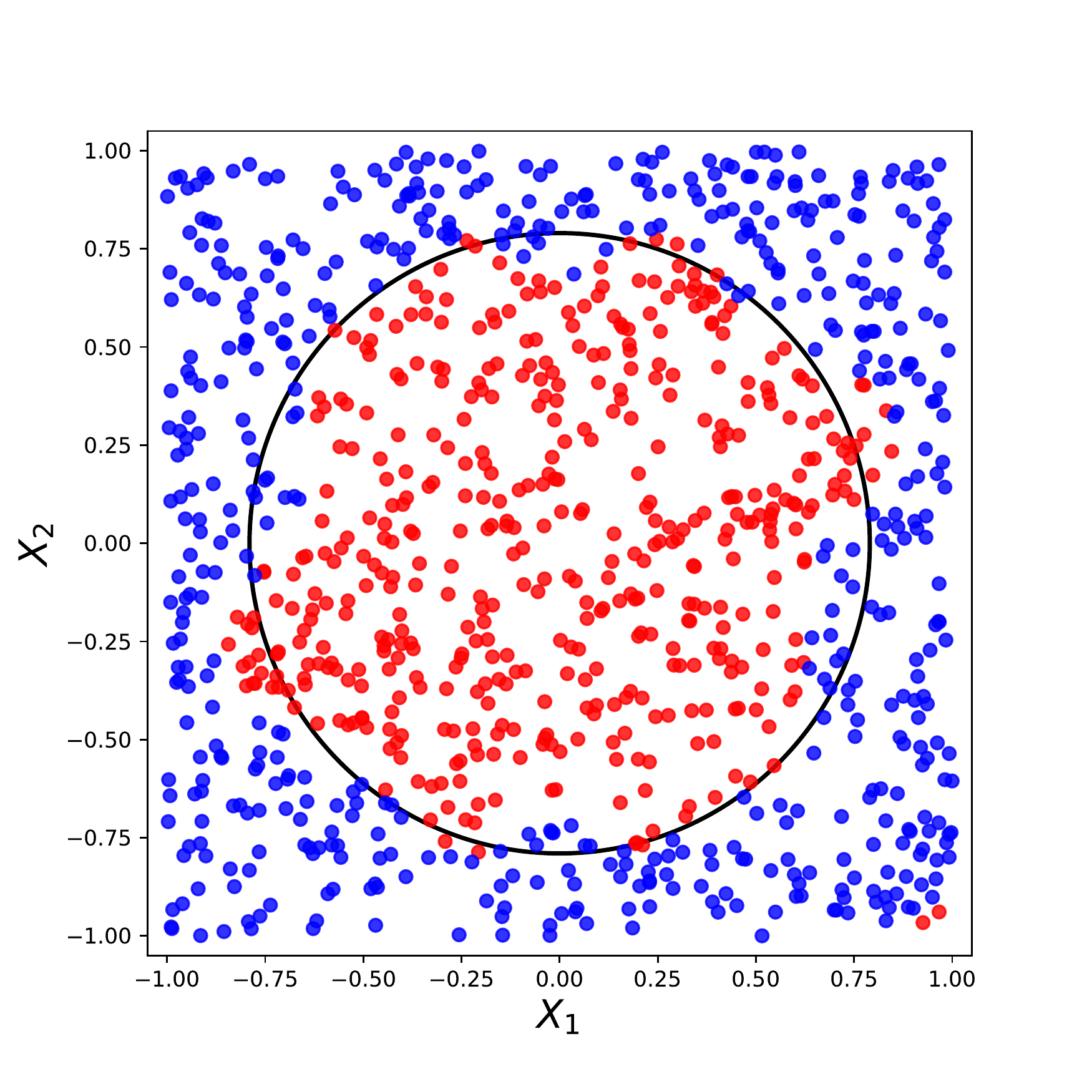}
\end{center}
\caption{Training and validation: based on the best parameters obtained after $12$ generations (training 2 of fig. 2c in the main text ) using the Genetic Algorithm (GA) optimizer, the classification results for a training dataset (left) comprising $250$ points and a test dataset (right) consisting of $1000$ points are depicted in this figure. The achieved accuracy of the classification on the test dataset is approximately $92.8\pm 1.8\%$, while the training accuracy reached $93.6\pm 1.8\%$ for a before-after comparison of the training procedure).}%
\label{fig:fig2}
\end{figure}
% %%%%%%%%

After the training is complete, a further step has been carried out in the quantum device by using the best optimal parameters obtained from training~$2$ (using GA optimizer) shown in fig.~(\ref{fig:fig2}) is the classification of $1000$ test data points. The reason to proceed in this way is to check the generalization capabilities of the quantum classifier without facing a too-costly optimization step. At the end of the execution, measurements are made to obtain the relative fidelity between the output state and all label states. The fidelity/accuracy obtained in this case (fig.~\ref{fig:fig2}, right) is about $92.8\%$ which is within the error-bar of the training accuracy (fig.~\ref{fig:fig2}, left), $93.6\pm1.8\%$. The error here refers to one standard deviation of $10$ repeated trials performed on the same data-set and it reflects the underlying systematic uncertainty leading to an uncertainty of the accuracy. Our experimental results confirm the trend seen by simulation, and the finite coherence and imperfections of the preparation of qubit do not seem to impact results significantly for the shallow circuit considered here.\\

Next, we introduce the experimental hardware stacks shown in fig.~(\ref{fig:full-stack}), starting with the quantum processor (QPU) and proceeding to the classical processor consisting of the middleware and the CPU.
\section{Experimental set-up}\label{sec:quantum-processor}
\begin{figure}
\begin{center}
\includegraphics[width=\linewidth]{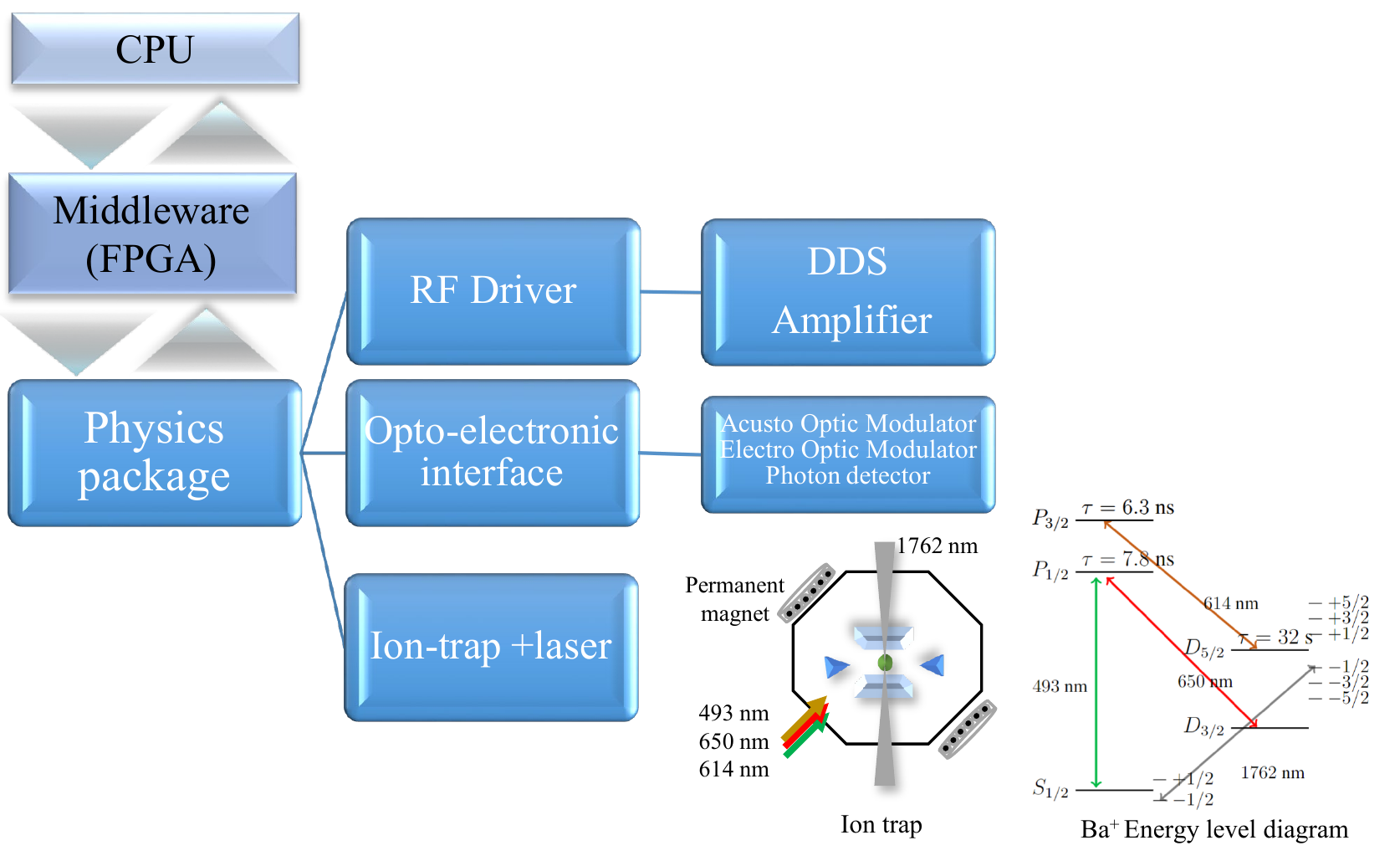}
\end{center}
\caption{The full-stack hybrid quantum-classical system comprises of three main layers: physics package or the quantum processing unit (QPU), middleware, and CPU. The QPU layer encompasses the ion trap setup along with lasers, the opto-electronic interface and the RF-drivers. The middleware is the interface between the QPU and the CPU and realized by a FPGA. The top-most layer is the CPU which implements the optimizer in a Linux environment using python script. In our ion trap setup, we specifically work with barium ions. The energy level diagram of barium ions, shown in the schematic, provides crucial information about the lasers involved in cooling, measurement, and gate implementation. Further details are discussed in the text.
}
\label{fig:full-stack}
\end{figure}

The full-stack quantum classifier hardware can be broken down into three functional stacks namely, the quantum processing unit~(QPU), the middleware and the classical processing unit~(CPU) as illustrated in fig.~(\ref{fig:full-stack}). The QPU and the middleware is very similar to our earlier work~\cite{Dutta-2022}, except that the classical processing unit~(CPU) is modified to efficiently perform training of the quantum-classical hybrid classifier for the task of binary classification. The QPU consists of three functional blocks namely, the ion trap and lasers, opto-electronic interface, and the RF drivers.\\
The ions are confined in a linear blade trap with axial confinement frequency $\sim 2\pi \times 0.5 $ MHz, and radial mode frequencies $\sim 2\pi \times 1.5 $ MHz. A magnetic field of $0.36$ mT provided via a low-temperature coefficient Sm$_2$Co$_{17}$ permanent magnets outside the vacuum chamber establishes the quantization direction, which is oriented $\sim 45\deg$ from the trap axis as shown in fig.~(\ref{fig:full-stack}). As in ref.~\cite{Dutta-2022, Dutta2020}, we choose ${\rm S}_{\frac{1}{2},-\frac{1}{2}}$ and ${\rm D}_{\frac{5}{2},-\frac{1}{2}}$  as the qubit's eigen states. This optical qubit transition frequency is weakly sensitive to magnetic field noise. We apply single qubit gates with a narrow-linewidth $1762~$nm laser locked to an ultra-stable cavity leading to a linewidth of $\approx 100~$Hz~\cite{Yum:17, Dutta2022j, ahmadi2024scalable}. Therefore, the significant contribution to de-phasing comes from magnetic field noise and residual time jitters of the gate pulses. The electro/acousto-optic (EO/AO) layer controls the phase, frequency and amplitude of the laser pulses during a gate implementation. Finally, the hardware control of the EOs and AOs is a combination of stable radio-frequency generators and RF amplifiers which in turn are controlled by the middleware consisting of field programmable gate arrays (FPGA) that produce the sequence of radio-frequency pulses as per the algorithmic instructions. The stable radio-frequency generators are made of direct digital synthesizers (DDS) based on analog device chip AD$9958$ capable of producing $20-250~$MHz with controllable frequency ($32$ bit resolution), phase ($16$ bit resolution) and amplitude ($10$ bit resolution). The FPGA, based on the Altera Cyclone V chip, controls the algorithmic time sequence as well as the measurement of the final state of the qubit.\\
%%%%%%%%%%%%%%%%
%%%%%%%%%%%%%%%

The quantum processor operates through a sequence comprising initialization, gate implementation, and quantum state determination. Prior to performing each training cycle, the qubit is first Doppler cooled to the Lamb-Dicke regime via a fast dipole transition (between S-P levels shown in the barium energy level diagram in fig.~(\ref{fig:full-stack}) at $493~$ nm, along with the simultaneous application of a re-pump laser (between D-P levels) at $650~$ nm. To cool all three normal modes of motion of the ion in the trap, care has been taken that the Doppler-cooling beam has sufficient overlap with all motional modes. In the following, we outline the specific steps of the QPU that are crucial for training, emphasizing their impact on error and fidelity.\\

\subsection{Initialization:} 

After the process of Doppler cooling, the ion population is distributed between the two magnetic sub-states of the ground state, ${\rm S}_{\frac{1}{2}}(m_j = +\frac{1}{2} $or $-\frac{1}{2})$, with a slightly higher probability of the ion being in the lower state due to different de-tunings with respect to the cooling light. To initialize the qubit in $m_j =-\frac{1}{2}$ state, the ion is optically pumped by applying resonant $1762~$nm light for about $200~\mu$s on the  ${\rm S}_{\frac{1}{2},+\frac{1}{2}}$ to ${\rm D}_{\frac{5}{2},-\frac{1}{2}}$ transition accompanied by a continuous $614~$nm light addressing ${\rm D}_{\frac{5}{2}}$ to ${\rm P}_{\frac{3}{2}}$. Additionally, $650~$nm light is simultaneously applied with the $614~$nm light to facilitate the depopulation of the ${\rm  D}_{\frac{3}{2}}$ state since decay from ${\rm P}_{\frac{3}{2}}$to ${\rm  D}_{\frac{3}{2}}$ is also allowed.\\ 
The state initialization process on the  ${\rm S}_{\frac{1}{2}}-{\rm  D}_{\frac{5}{2}}$ transition is somewhat more complex than using circularly polarized light on the ${\rm S}_{\frac{1}{2}}-{\rm  P}_{\frac{1}{2}}$ or ${\rm  P}_{\frac{3}{2}}$ transition. However, it offers a distinct advantage. On the ${\rm S}_{\frac{1}{2}}-{\rm  D}_{\frac{5}{2}}$ transition, the magnetic sub-levels can be distinguished by their distinct frequencies. This makes it possible to achieve high-fidelity initialization without the need to be concerned about geometric alignment with respect to the magnetic field, which could otherwise introduce unwanted polarization components. With this procedure, we get pumping efficiencies exceeding $98.7\pm0.5\%$ for a single $^{138} {\rm Ba}^+$ ion. Once the qubit is initialized, any single qubit rotational gate is implemented by resonantly driving the qubit with full control over the laser phase, power and laser on time. The key parameters of the trapped-ion system used in this experiment are the qubit coherence time of 5 ms and a Rabi $\pi$-time of 12 µs, with a gate fidelity of $98.7$$\%$.
\\
\subsection{Quantum gate implementation:} 

The quantum circuits designed for the training process of the classification task on the quantum processor are depicted at the bottom of the schematic in fig.~(\ref{fig:training}). The quantum circuit is defined as a series of non-commutating rotational gates as elucidated in reference~\cite{Dutta-2022, perez2020data}. Each layer of circuit consists of two gates, and in our implementation, we utilize a total of $4~$layers as it was found to be optimal~\cite{Dutta-2022}. Depending on the data to be uploaded, single qubit gates are implemented by resonantly driving the qubit transition between $\ket{0}$ and $\ket{1}$ with well-controlled phase and operation time, while maintaining constant laser frequency and power. The schematic description of the training algorithm employed in this work is illustrated in fig.~(\ref{fig:training}). The optimization process continues until optimal values of $\bar{\Theta}$ are obtained, ensuring a classification accuracy exceeding $90\%$. The crucial step after the implementation of the quantum gates as per the algorithm is the quantum state determination which is discussed in the following.\\

\subsection{State determination:} 

At the end of each measurement, the final state $\ket{\phi}$'s projection on to the label state $\ket{y}$ is measured, and the result is used to compute the cost function (CE) that quantifies the error made in the classification of the training set. In our experimental setup, the state projection is carried out at the end of each experimental sequence using a Photo-Multiplier Tube (PMT) to register the incoming photon number emitted by the ion within a predefined time window. For a time window of $2$ ms, the fluorescence lasers at $493~$nm (along with $650~$nm) are switched on and all photons detected perpendicular to the laser propagation direction during that time are added up by a digital counter in the middleware. If the number of detected photons is above a predefined threshold, we assign the ${\rm  S}_{\frac{1}{2}}$ state to the result, if it is below we assign the ${\rm  D}_{\frac{5}{2}}$ state. The number of events above/below the threshold relative to the total amount of repetitions determines the probabilities $p1$ and $p0$ (equation 14) corresponding to population of the two states of the qubit. A typical histogram of one of our measurements, depicting the counts acquired in $2$ milliseconds with $200$ repetitions, is illustrated in fig.~(\ref{fig:single-state}). After collecting the fluorescence for an integration time of $2$ ms, we use aforesaid threshold to determine the state of the ion, discriminating the quantum state with more than $99\% $ accuracy.  
Achieving high detection fidelity comes at the cost of long state detection time leading to errors. These errors include finite spontaneous decay probability of the D$_{5/2}$ state during the detection period, dark counts of the photo-detector, scattered photons from the laser beam, and Poisson statistics of the bright state. Since, an ideal classifier should operate in a noisy environment, next, we will analyse the classification error rather than the systematic errors of individual gates.
\subsection{Error sources:}

Some errors in quantum computers are coherent, implying that imperfect operations remain Hermitian, and applying their inverse restores the system to its previous state. Examples of such errors encompass instances where qubits experience rotations with inaccurately determined Rabi or resonant frequencies. The impact of these issues is reflected in the fig.~(\ref{fig:training-dataset}), yet these errors can be mitigated through the implementation of enhanced control schemes. Improvements can be realized through optimized beam delivery or the application of beam pulse shaping techniques. In sec.~(\ref{error-analysis}), we provide an in-depth account of our efforts to rectify these errors and detail the shortcomings.

The fidelity of quantum computation faces limitations due to experimental noise, causing the system to deviate from its ideal evolution. This deviation can originate from various sources beyond SPAM errors, which are also examined in detail.
%%%%%%%%%%%%%%%%%
\begin{figure}
\begin{center}
    \includegraphics[width=\linewidth]{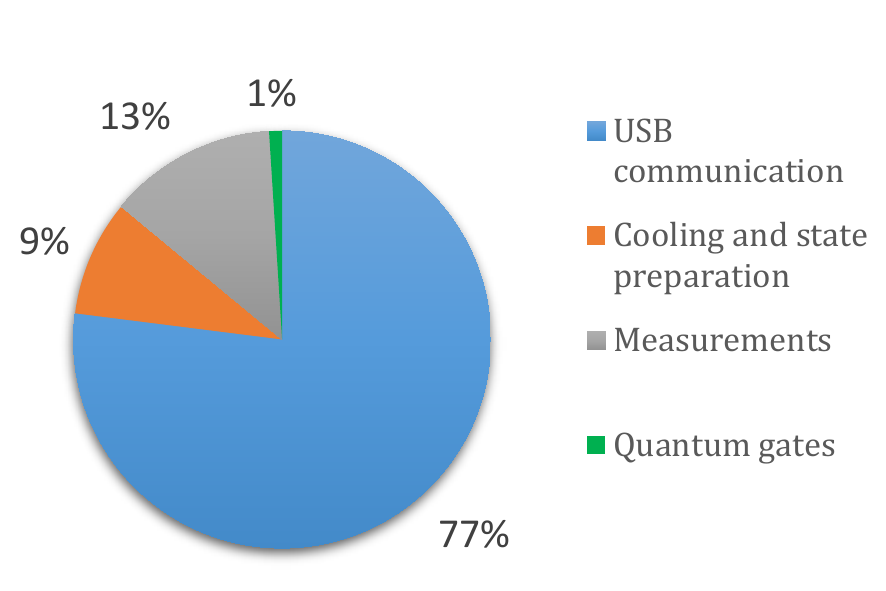}
\end{center}
\caption{Time Budget: the time required to complete the training of one generation for a training dataset of $250$ data points, with a set of $50$ individuals, is approximately $330$~min in our current setup. Each data point in the training dataset is iterated $150$ times. The distribution of time for each step is depicted in the PI-chart. A significant portion of the time is allocated to USB serial communication, which mainly involves loading data onto the Field-Programmable Gate Array (FPGA), loading data onto the Direct Digital Synthesis (DDS), and receiving data from the FPGA. The remaining time is dedicated to tasks related to cooling, measurement, and gate implementation.}
\label{fig:time}
\end{figure}
%%%%%%%%%%%
\subsection{Time Budget:} 
Training quantum computers can be a time-consuming process, especially when dealing with a quantum machine that is noisy. We are utilizing a hybrid quantum-classical approach where classical computations are used in conjunction with quantum computations to speed up training and optimization processes, still, the time it takes to train our QPU for each generation is about $330$~minutes, each generation has a set of $50$ individuals. So, in our case, the current bottleneck is the time spent on data transfer between QPU and CPU via USB is significantly high compared to other aspects of quantum computing, such as quantum gate implementation, cooling, preparation, and measurement time. The details of the time budget for our setup are presented in fig.~(\ref{fig:time}). This can be mitigated by efficient design of the middleware. The data re-uploading algorithm avoids directly loading classical data to the quantum states thereby saving the circuit depth. However, it needs commensurate classical memory in the middleware to continually feed the classical data. Next, we discuss the methods used to improve the efficiency of training of the classifier on a hybrid system.

%%%%%%%%%%%%%%%%%%%%%%%%%%%%%%%%%%%%%%%%%%%%%%%%%%%%%%%% End of section experimental setup%%%%%%%%%%%
\section{Efficient training of NISQ classifier}
\label{sec:opt}
In the NISQ era, the QPU is noisy but the classical interface is well-developed. 
\subsection{Choice of kernel}\label{sec:kernel}

\begin{figure}[t]
\begin{center}
\subfigure[\hspace{0mm} ]{
        \includegraphics[width=0.4\linewidth]{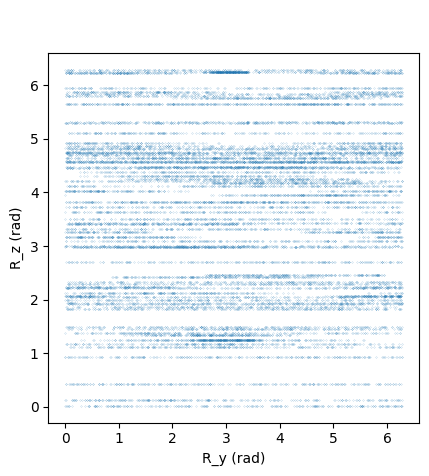}
      }
    \subfigure[\hspace{0mm} ]{
        \includegraphics[width=0.4\linewidth]{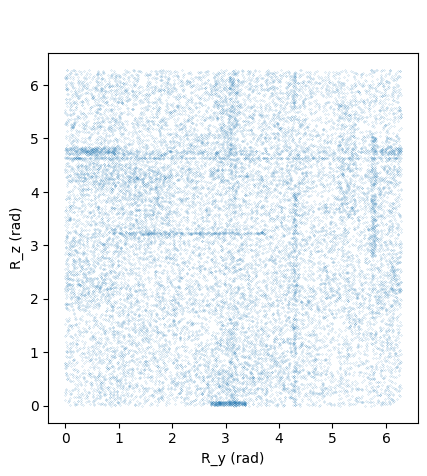}
      }
    \subfigure[\hspace{0mm} ]{
        \includegraphics[width=0.4\linewidth]{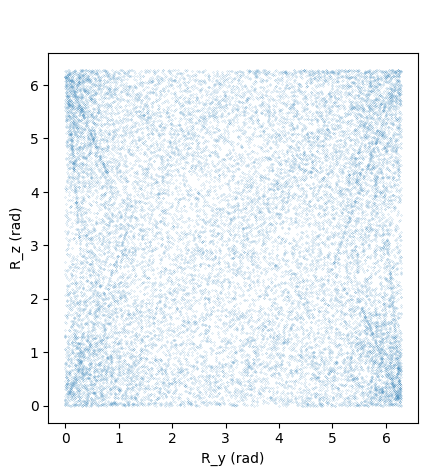}
      }
    \subfigure[\hspace{0mm} ]{
        \includegraphics[width=0.4\linewidth]{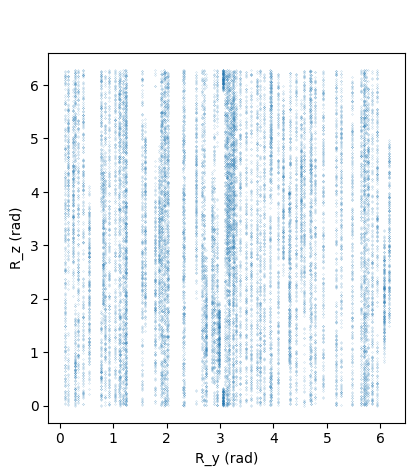}
      }
\end{center}
\caption{Aggregate rotational values of 20 randomly chosen sets of parameters using different Ansatz. (a) Ansatz 2A. (b) Ansatz 2B. (c) Ansatz 2C. (d) Ansatz 2D. }
    \label{fig:analysis-of-ansatz}
\end{figure}

The kernel function plays a crucial role in the performance of a classifier as pointed in \cite{steinwart2001influence} for support vector machine~(SVM) and classical neural networks (CNN). In the data re-uploading setting, the kernel also plays an important role, both in exploration and exploitation of the parameter space. There are four possible kernels applicable to our classifier by considering only linear mapping; these are:\\

\begin{align*}
    \text{Ansatz-2A: } & U_i (\bar{\theta}, \bar{x}) = R_z (\theta_3) R_y (\theta_0 x_0 + \theta_1 x_1 + \theta_2) \\
    \text{Ansatz-2B: } & U_i (\bar{\theta}, \bar{x}) = R_z (\theta_2 x_1 + \theta_3) R_y (\theta_0 x_0 + \theta_1) \\
    \text{Ansatz-2C: } & U_i (\bar{\theta}, \bar{x}) = R_z (\theta_2 x_0 + \theta_3 x_1) R_y (\theta_0 x_0 + \theta_1 x_1) \\
    \text{Ansatz-2D: } & U_i (\bar{\theta}, \bar{x}) = R_z (\theta_1 x_0 + \theta_2 x_1 + \theta_3) R_y (\theta_0). \\
\end{align*}
We note that the number $2$ in the above alternative kernel label refer to the dimension of the input size, a convention that was introduced in \cite{perez2020data}.

To compare the influence of the four choices of kernel, a set of $250$ training data points from the binary classification problem are fixed. In fig.~(\ref{fig:analysis-of-ansatz}), we plot the aggregate $R_y$ and $R_z$ rotational values against each other. These plots illustrate the exploration of a part of the feature space based on the kernel used. As we can see, Ansatz 2A and 2D have the least amount of spread because their degree of freedom for $R_z$ and $R_y$, respectively, are limited. On the other hand, Ansatz 2B and 2C's spreads are more comparable with 2B having slightly more clustering around certain regions. Given the results of the above analysis, we chose Ansatz 2C considering these analysis above during the training of our quantum classifier.\\

As a precursor to our training algorithm, we note that the particular choice of Ansatz has less impact for gradient-based optimization methods but affect both the rate of convergence and final convergence value for genetic-based optimization methods.\\
%%%%%%%%%%%%%%%%%%%%%%
\subsection{Optimization Strategies}

With the kernel function fixed, we are ready to proceed to the last two pieces of our algorithm from \textit{Step 8} of fig.~(\ref{fig:training}): the \textit{cost function} and \textit{classical optimizer}. The particular choice of these pieces both determines the rate of convergence and the converged value. Before we dive in, we will concretely define our measurement function. Given a kernel with $L$ layers, parameters $\bar{\theta}$, a data point $\bar{x}$ and a label $y$, the measurement function returns the expectation value of the final state projected onto the label state:

\begin{align}
    M(\bar{\theta}, \bar{x}, y) = \bigl| \bra{y} \prod_{l=0}^L U_l(\bar{\theta}, \bar{x})\ket{0} \bigr|^2
\end{align}

We now examine various choices of cost functions and optimizer methods.

\subsubsection{Cost Functions}\label{sec:cost-functions}

\begin{figure}[h]
\begin{center}
\includegraphics[width=\linewidth]{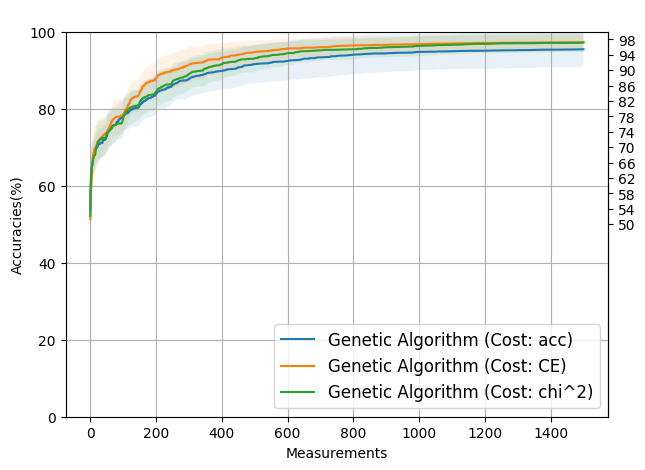} 
\end{center}
\caption{Simulation results of highest cumulative GA accuracy with three different metrics. Cross-entropy loss gives best rate of convergence. Both cross-entropy and chi squared loss shows comparable final accuracy. Using accuracy as loss shows both low converged value and slower rate of convergence.}
    \label{fig:cost-functions}
\end{figure}

A natural way to define the cost function is the accuracy of our quantum classifier since we are dealing with a classification problem. Because this is a binary classification task, a data point is considered \textit{correctly} classified if the projection onto the label state has probability higher than $0.5$. Given a parameter $\bar{\theta}$, a set of data points $\bar{\mathbf{x}}$ and labels $\mathbf{y}$, the accuracy function is

\begin{align}
    acc(\bar{\theta}, \bar{\mathbf{x}}, \mathbf{y}) = - \frac{1}{n} \sum_{i=1}^n \mathbbm{1}_{M(\bar{\theta}, \bar{x}_i, y_i) > 0.5}
\label{eq-cf1}
\end{align}

In the equation above, $\mathbbm{1}$ represents the indicator function. A noticeable caveat to this choice is that as the accuracy improvements it gives \textit{sparse} signals. That is, a classifier that correctly classifies all data points with $100\%$ confidence is indistinguishable from one that merely gives $51\%$ confidence under this metric. Indeed, it is desirable to have cost functions that are \textit{continuous} and \textit{differentiable} in classical neural networks. We now consider a common cost function that is often used in classical machine learning: \textit{cross-entropy (CE) loss}.

\begin{align}
    CE(\bar{\theta}, \bar{\mathbf{x}}, \mathbf{y}) = - \frac{1}{n} \sum_{i=1}^n \mathbbm{1}_{M(\bar{\theta}, \bar{x}_i, y_i) > 0.5} \log{M(\bar{\theta}, \bar{x}_i, y_i)}
    \label{eq-cf2}
\end{align}

We note that the CE loss attaches a log probability to the indicator function, which helps join the discontinuity when the probability shifts from below 0.5 to above 0.5. Lastly, we also consider a third cost function from \cite{perez2020data}, \textit{chi squared ($\chi^2$) loss}.

\begin{align}
    \chi^2(\bar{\theta}, \bar{\mathbf{x}}, \mathbf{y}) = \frac{1}{n} \sum_{i=1}^n (1 - M(\bar{\theta}, \bar{x}_i, y_i))^2
\label{eq-cf3}
\end{align}

Similar to the CE loss, $\chi^2$ loss is also continuous and differentiable. However, we note two difference between the two loss functions: 1) the $\chi^2$ loss converges faster to 0 as the correct classification probability gets close to 1; and 2) the $\chi^2$ loss is capped at 1 for incorrect classifications whereas the CE loss is not bounded above. In practice, both $\chi^2$ loss and CE loss lead to comparable performance as we can see from our simulation results in fig.~(\ref{fig:cost-functions}). However, the rate at which CE ({\color{orange}$-$}) responds to the parameter changes is higher and hence crucial to achieve optimal value in shorter time. In the following, we discuss the basics of each of the classical optimizer that are compared in the main article.
%%%%%%%%%%%%%%%%%%%%%
\subsubsection{Genetic Optimizer}\label{sec:genetic-optimizer}

The Genetic Algorithm (GA) is a family of optimization algorithms inspired by the process of natural selection, emphasizing the survival of the fittest principle. According to a recent survey, the general form of GA commences with the initialization of a population (Y) comprising n randomly generated chromosomes. Following this, the fitness of each chromosome is evaluated, forming the basis for the selection process wherein two chromosomes, denoted as C1 and C2, are chosen based on their respective fitness values \cite{katoch2021review}. 

To thoroughly investigate these dimensions, we will first present a high-level abstract algorithm tailored at the quantum classifier problem, and then provide an analysis of each of these optimization dimensions.

In the setting of the quantum classifier described above, we may define a skeleton GA algorithm algo.~(\ref{algo:genetic}) by letting $\theta \in \mathbb{R}$ be an individual gene, and a vector of such genes, $\bar{\theta} \in \mathbb{R}^n$, be a chromosome. Our initial population will be comprised of a set of randomly generated chromosomes. The population is then evaluated and carries out reproduction according some fitness function $f$. We will also discuss the choice of $f$ in detail.

\begin{figure}[h]
\begin{center}
    \includegraphics[width=\linewidth]{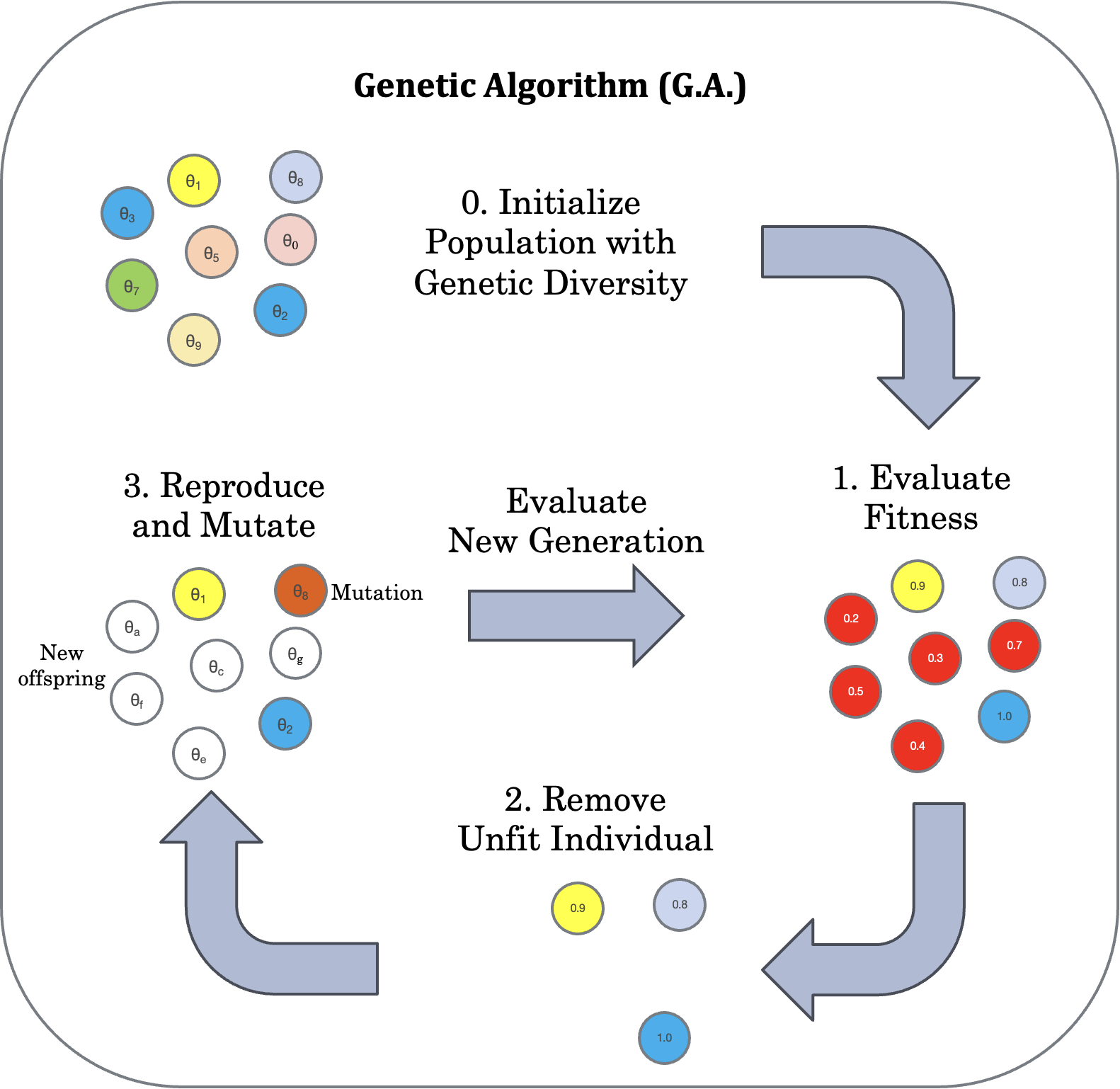} 
\end{center}
\caption{Life-cycle of genetic algorithm. This serves as a visual illustrate to Algorithm \ref{algo:genetic}.}
    \label{fig:genetic-algorithm}
\end{figure}

\begin{algorithm}
    \caption{Genetic Algorithm for Quantum Classifier}
    \label{algo:genetic}
    \KwData{Population Size $\leftarrow$ S}
    \KwData{Convergence function, $Converged(\ldots)$, returns True if stopping criteria are reached}
    \KwResult{Best solution, $\bar{\theta}_{\text{best}}$}
    
    \SetAlgoNlRelativeSize{0}
    \SetAlgoNlRelativeSize{-1}
    \SetAlgoNlRelativeSize{1}
    
    Generate initial population of S chromosomes $\bar{\theta}_1, \bar{\theta}_2, \ldots, \bar{\theta}_S$\;
    Initialize iteration counter $t=0$\;
    Compute the fitness value of each chromosome, $f(\theta_i)$\;
    
    \While{not $Converged(t, f(\theta_1), f(\theta_2), \ldots, f(\theta_S))$}{
        Randomly select a few chromosomes to allow for survival to the next iteration\;
        Select one or more chromosomes from the current population based on fitness and hyper-parameters\;
        Apply crossover operation\;
        Apply mutation\;
        Replace the old population with the newly generated population\;
        Update $t\leftarrow t+1$\;
    }
    
    \Return $\bar{\theta}$ with the highest fitness value\;
\end{algorithm}

Like other machine learning algorithms, GA has various hyper-parameters that we need to tune. Here are a few common ones we investigated.

\begin{figure}[h]
\begin{center}
    \subfigure[\hspace{0mm} ]{
        \includegraphics[width=0.45\linewidth]{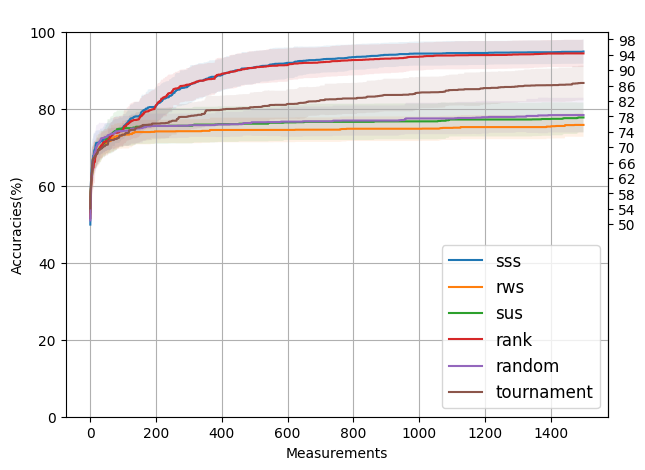}
      }
    \subfigure[\hspace{0mm} ]{
        \includegraphics[width=0.45\linewidth]{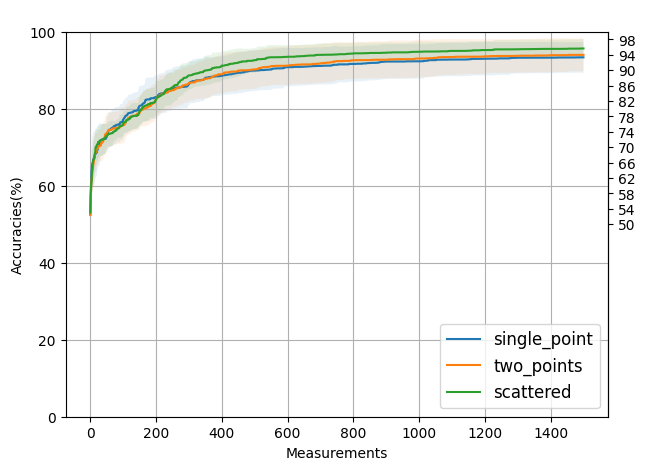}
      }
    \subfigure[\hspace{0mm} ]{
        \includegraphics[width=0.45\linewidth]{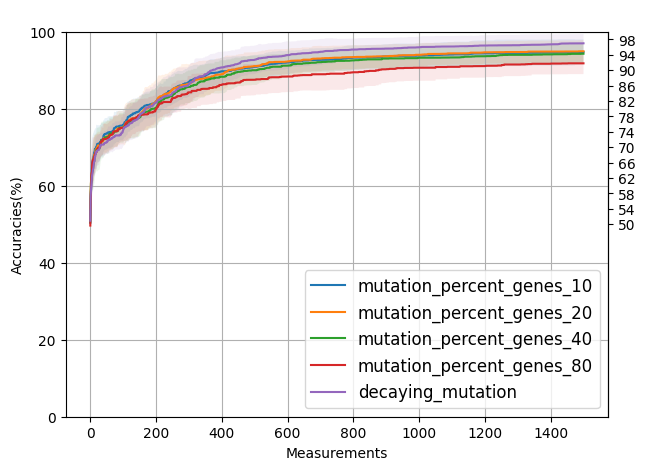}
      }
    \subfigure[\hspace{0mm} ]{
        \includegraphics[width=0.45\linewidth]{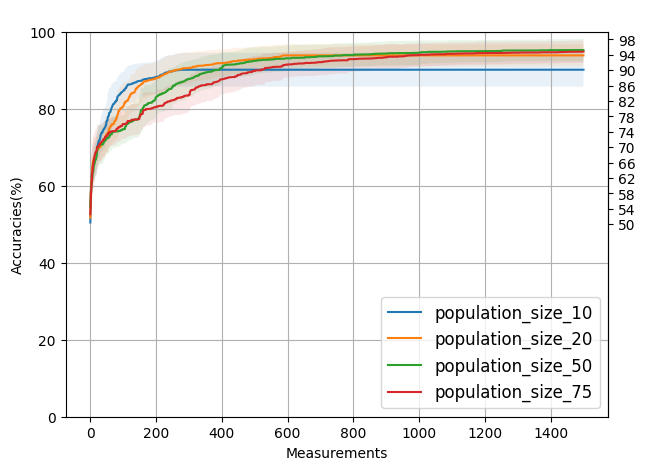}
      }
\end{center}
\caption{Simulation results on various GA hyper-parameters. For all plots, cumulative highest accuracy is plotted against measurements. (a): comparison of GA with steady state selection (sss), roulette wheel selection (rws), stochastic universal selection (sus), rank selection (rank), random selection (random), and tournament selection (tournament). (b): comparison of GA with single point, two points, and scattered crossover. (c): comparison of GA with 10\%, 20\%, 40\%, 80\% fixed mutation rates, and decaying mutation. (d): comparison of GA with 10, 20, 50, and 75 population pool.}
    \label{fig:analysis-of-ga-hyperparams}
\end{figure}

\begin{enumerate}
    \item Encoding Scheme of individual genes:\\
    \begin{itemize}
        \item The authors in \cite{katoch2021review} list several commonly used encoding schemes for translating a problem into a GA-applicable one. They also suggest that value-based encoding schemes are usually preferred in neural networks for finding the optimal weights. Our problem is closely related to optimizing a neural network and we used value-based encoding scheme.
    \end{itemize}
    \item Selection criteria for chromosome reproduction:\\
    \begin{itemize}
        \item The selection function in GA determines whether a particular chromosome will be selected to reproduce based on its fitness value. For the quantum classifier, we experimented with several selection techniques in simulation. We found that Steady-State Selection (SSS) is a superior selection method (see fig. (\ref{fig:analysis-of-ga-hyperparams})) and this is what we used in practice. Other selection functions often lead to premature convergence in our problem setting.
        \item We also note here that the selection function acts as a convergence pressure for the GA. And this can be thought of as a knob that we can tune to trade-off exploration and exploitation to converge to the best possible value given a training iteration constraint.
    \end{itemize}
    \item Crossover type to decide how off-springs inherit parents' genetic information:\\
    \begin{itemize}
        \item Crossover type determines how off-springs are produced using the genetic information from 2 or more parents. We compared one-point, two-point and scattered crossover when 2 parents are selected for mating. We find that all three crossover types offer comparable performance in our problem setting. And unlike selection function, no crossover type suffers significantly from premature convergence and scattered crossover seems to give slightly better performance (see fig.~(\ref{fig:analysis-of-ga-hyperparams})). This is what we used for our experiment.
    \end{itemize}
    \item Mutation function to decide how often and how much mutations occur:\\
    \begin{itemize}
        \item In our simulation runs, we see various mutation probabilities significantly impacts convergence. With fixed mutation probability, GA performs the best with a mutation rate of 20\% and gradually degrades with higher mutation values. This is most likely caused by an over-emphasis on the exploration of our parameter space rather than exploiting what we have learned already.
        \item In classical deep learning, adaptive learning rate and weight decay are often employed as explicit knobs to decide how the rate of exploration-exploitation trade-off changes \cite{smith2018disciplined}. Borrowing this idea, we also introduced the concept of decaying mutation rate: $\mathbbm{1}_{p(\text{maskbase}^t)} * \delta * t^{\text{scale}}$. Here maskbase and scale are both hyper-parameters and $t$ is the iteration number. $\delta$ is a uniform random variable between -0.5 and 0.5. We observe that this leads to better performance when compared to using a constant mutation rate as used for fig.~(\ref{fig:analysis-of-ga-hyperparams}).
    \end{itemize}
    \item Size of the population pool:\\
    \begin{itemize}
        \item Population size is the most important factor in determining how well a GA converges. From fig.~(\ref{fig:analysis-of-ga-hyperparams}), we can see that lower population tends to converge early to unfavorable local minima and higher population explores for longer but eventually converges to better minima. We note that unlike the previous hyper-parameters, population size is directly related to amount of resource required. In practice, resource is scarce and our choice of population size was $50$ during our GA experiments.
    \end{itemize}
\end{enumerate}
% \subsection{Genetic Optimizer}

\begin{figure*}[]
\begin{center}
\label{fig:fig4c}
        \includegraphics[width=\textwidth]{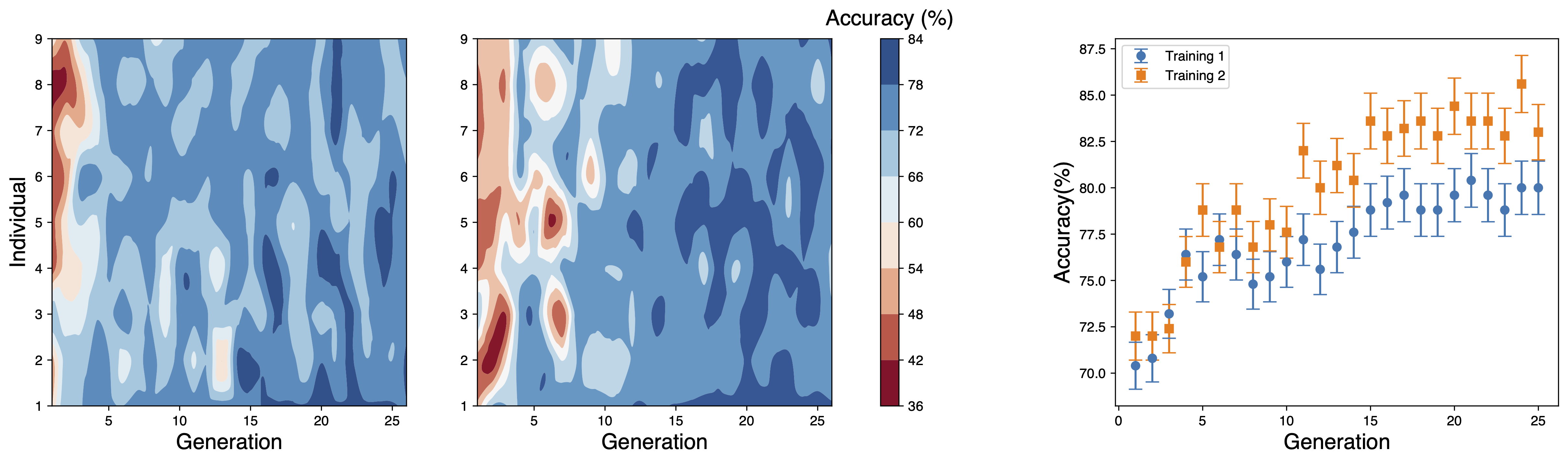}
\end{center}
\caption{\textit{Left}: Training run 1. \textit{Center}: Training run 2. \textit{Right}: Search for optimal parameters for classification task using Genetic algorithm. The ion trap-based QPU is used for training on $250$ random data points. The depth of the circuit is kept fixed to 4 layers as it is sufficient to classify the current problem.  For this training, a set of 9 individuals have been used for training and the data has been uploaded using ansatz 2A.
}
    \label{fig:ga-9-pop}
\end{figure*}

In addition to the results with genetic optimizers in the main article, we also include a couple of experimental runs with population sizes of $9$. As we can see from fig.~( \ref{fig:ga-9-pop}), this setup suffers from premature convergence due to low population size as shown in our simulation runs in sec.~(\ref{sec:genetic-optimizer}). Next, we delve deeper into the use of gradient-based optimizer in NISQ devices, their pros and cons. 

\subsubsection{Gradient-based Optimizer}
\label{gradient}
\begin{algorithm}
    \caption{Gradient Optimization for Quantum Classifier}
    \label{algo:gradient}
    \KwData{Initial parameter $\bar{\theta}$}
    \KwData{Gradient Optimizer Function, \textit{Opt}, returns \textbf{False} if converged}
    \KwResult{Best parameter $\bar{\theta}_{\text{best}}$}
    
    \SetAlgoNlRelativeSize{0}
    \SetAlgoNlRelativeSize{-1}
    
    Initialize iteration counter $t=0$\;
    Function to compute metric of parameter, $f(\bar{\theta})$\;
    Initialize $\bar{\theta}_{\text{best}}=\bar{\theta}$\;
    Initialize $\bar{\theta}_{\text{new}}=\bar{0}$\;
    
    \While{$Opt(t, \bar{\theta}, f, \bar{\theta}_{\text{new}})$}{
        Conditionally update best parameter, $\bar{\theta}_{\text{best}}$, to $\bar{\theta}_{\text{new}}$\;
        Update $\bar{\theta}\leftarrow\bar{\theta}_{\text{new}}$\;
        Update $t\leftarrow t+1$\;
    }
    
    \Return $\bar{\theta}_{\text{best}}$\;
\end{algorithm}

Another class of popular algorithm is gradient-based methods, which is an iterative optimization method that leverages first and higher order derivatives. These methods are very popular in classical machine learning algorithms as they are the $\textit{de\ facto}$ method used in optimizing deep neural networks. We explore two gradient-based methods in this paper. First is Stochastic Gradient Descent (SGD), which is a very popular method in the classical deep learning community. Second is a quasi-newton method, Broyden-Fletcher-Goldfarb-Shanno (BFGS) algorithm.

\paragraph{Stochastic Gradient Descent}

Stochastic Gradient Descent (SGD) is a widely used optimization technique, especially prevalent in the training of machine learning models. It is an iterative method for optimizing an objective function with suitable smoothness properties (e.g., differentiable or sub-differentiable). The core principle of SGD is to perform the optimization by updating the parameters incrementally, using a subset of training samples at each step. This is in contrast to traditional gradient descent, which uses the entire data set to compute the gradient at each iteration. Mathematically, the update rule of SGD in the quantum classifier setting can be defined as:

\begin{align*}
    \bar{\theta}_{t+1} = \bar{\theta}_t - \text{lr} \frac{1}{\text{Batchsize}} \sum_{\bar{x} \in \text{mini-batch}} \nabla f(\bar{\theta}_t, \bar{x})
\end{align*}

where:
\begin{itemize}
  \item \(\bar{\theta}_t\) represents the parameter vector at iteration \(t\).
  \item lr is the learning rate, a hyper-parameter that determines the step size at each iteration.
  \item \(f(\bar{\theta}_t, \bar{x})\) is the average loss function computed for the mini-batch.
  \item \(\nabla f(\bar{\theta}_t, \bar{x})\) denotes the gradient of the loss function with respect to the parameters.
\end{itemize}

By utilizing only a few training examples at each step, SGD significantly speeds up the optimization process, albeit at the cost of a more noisy convergence path. In simulation, we have found that SGD's convergence is significantly impacted by this noise and decided not to implement it for the experiment, which is an even noisier environment. We also found that even when we take the batchsize to be the entire training set, the rate of convergence is inferior to other optimizers and decided against running it in our experimental setup.

\paragraph{Quasi-Newton Method}

The Broyden–Fletcher–Goldfarb–Shanno (BFGS) method is a prominent iterative method for solving unconstrained nonlinear optimization problems. It uses second derivative to help guide the optimization path and takes a better convergence path when compared to gradient descent or SGD methods. It also belongs to quasi-Newton methods, a category of popular optimization techniques that approximate the Hessian matrix of second derivatives, which is a lot less expensive than computing the Hessian directly. In the context of optimization, the BFGS method seeks to find the minimum of a function \( f \), where the update rule for each iteration \( k \) is given by:

\[
\bar{\theta}_{k+1} = \bar{\theta}_k - \alpha_k B_k^{-1} \nabla f(\bar{\theta}_k)
\]

The BFGS method approximates \( B_k \), the inverse of the Hessian matrix, using the following update rule:

\[
B_{k+1} = B_k - \frac{B_k y_k y_k^T B_k}{y_k^T B_k y_k} + \frac{s_k s_k^T}{y_k^T s_k}
\]

where:
\begin{itemize}
  \item \( \bar{\theta}_k \) represents the parameter vector at iteration \( k \).
  \item \( \alpha_k \) is a scalar that typically satisfies the Wolfe conditions.
  \item \( \nabla f(\bar{\theta}_k) \) is the gradient of the function \( f \) at \( \bar{\theta}_k \).
  \item \( s_k = \bar{\theta}_{k+1} - \bar{\theta}_k \) is the difference in consecutive parameter vectors.
  \item \( y_k = \nabla f(\bar{\theta}_{k+1}) - \nabla f(\bar{\theta}_k) \) is the difference in consecutive gradients.
  \item \( B_k \) is the approximation to the inverse Hessian matrix at iteration \( k \).
\end{itemize}

\paragraph{Gradient Estimation}

The BFGS method is renowned for its superior convergence properties. In order to leverage this algorithm, it remains to estimate the gradient at each desired $\bar{\theta}$ values. For classical machine learning problems, BFGS-based methods usually approximate derivatives using the finite difference approach. Ideally a step-size below 0.1 is needed to guarantee goodness of convergence. However, this step-size is hard to achieve in practice due to random noises in our system. We also attempted a different technique called $\textit{parameter shift}$ to combat this issue \cite{crooks2019gradients}. This technique allows for the estimation of gradient at a particular parameter using arbitrary shifts in two directions. We fixed a shift size of $\frac{\pi}{2}$ in order to maximally leverage this technique and reduce the effect of random noise. We defer readers to sec.~(\ref{error-analysis}) for detailed experimental analysis.
%%%%%%%%%%%%%%%%%%%%%%%

\subsubsection{Classical processor}

The rising popularity of deep learning lies in its ability to solve more and more problems better than humans. To this end, we also attempted training the quantum classifier using a classical reinforcement learning (RL) agent. We formulated our problem as an RL problem and used the DQN algorithm~\cite{mnih2015human}, with the following problem setup:

\begin{itemize}
    \item[State:] A vector of our classifier's parameter values.
    \item[Action:] Increment, decrement, tiny increment, tiny decrement for each parameter. In addition, we have a reset action that either randomizes our parameter values for random initial value setup, or resets the parameter back to the initial value for fixed initial value setup. We note that these increments and decrements are fixed with some uniform noise.
    \item[Next State:] The state we reach by taking an action.
    \item[Reward:] Let $A_i$ denote the accuracy at step i. The reward at step t is defined as $\max(0, \frac{A_t - \max_{j<t}A_j}{100-A_0})$.
\end{itemize}

%%%%%%%%%%%%%%%%%%%%%%%
\begin{figure*}
\begin{center}
\includegraphics[width=\linewidth]{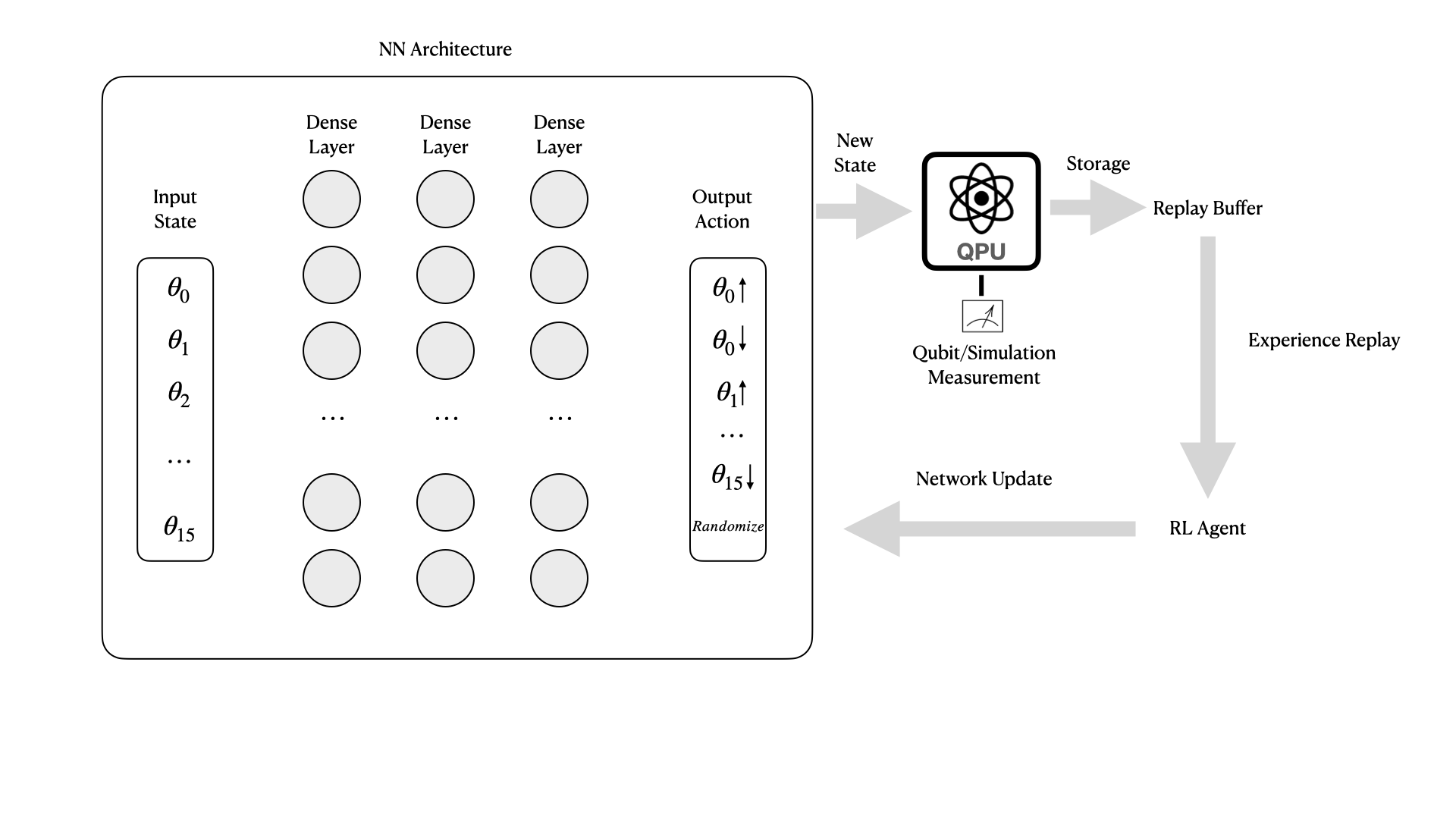}
\end{center}
\vspace{-20 mm}
\caption{Deep Reinforcement Learning (RL) Architecture used as a training agent for the quantum classifier. The values of our trainable parameters are the states of our RL problem. There are 4n+1 actions available for the agent to select from. For each parameter, we have 4 actions: increment, decrement, tiny increment, and tiny decrement. Finally we have a \textit{randomize} action that either picks a completely random value, or reset to the initial value depending on the training paradigm. The new state is measured on the simulation, and the reward (increase in accuracy/cross-entropy is stored in a replay buffer. The agent then randomly samples mini-batches of tuples (state, action, next state, reward) for training. The network is updated after each training step.
}
\label{fig:rl-architecture}
\end{figure*}

\begin{figure}
\begin{center}
    \includegraphics[width=\linewidth]{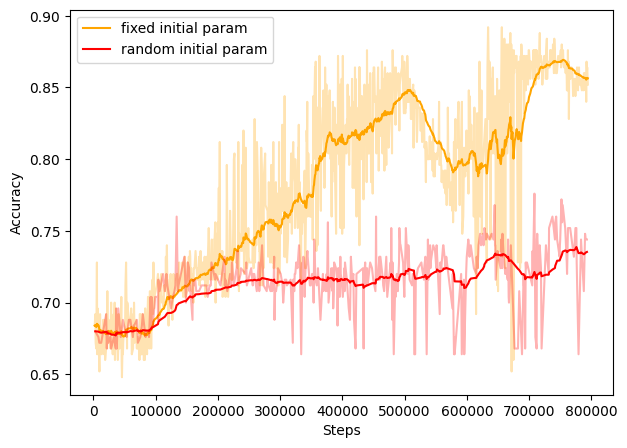}
\end{center}
\caption{Each step represents an agent taking 500 actions. Highest training accuracy achieved within 500 iterations is recorded as a single data point. The plot shows two training runs each with around 800,000 training steps. Faint lines represent the raw data and solid lines represent exponentially smoothed values with a smoothing factor of 0.95.
}
    \label{fig:rl-training}
\end{figure}

With the above reward definition, we note that the agent is rewarded for finding higher accuracy values, but will not be penalized if it has to temporarily go through regions with low accuracy values. The overall architecture is show in fig.~(\ref{fig:rl-architecture}). We used a densely connected neural network with $3$ hidden layers of size $512$. We used ReLU as our activation layers. The input size of our problem is vectors of length $16$ and the action space is of size $65$, using the above definition.

We attempted to train two different agents under this setup. A \textit{random initial param} agent that always starts or resets with some randomized parameter values. A \textit{fixed initial param} agent that picks a vector of $16$ random parameter values at step $0$ and reuses it for the entire training process. The agent is allowed to take $500$ actions and its total reward is the sum of discounted rewards using under the standard DQN paradigm. Our replay buffer has up to $1,000,000$ entries and our discount factor is $0.99$. We used a learning rate of $1^{-4}$  and mini-batch size of $32$. The exploration rate starts at $1$ and decreases to $0.01$ after $250,000$ steps.

We can see from fig~(\ref{fig:rl-training}) that the random agent learns very limited amount of information while the fixed agent reaches accuracy values above $85\%$ towards the end. We believe this is due to the problem landscape as illustrated in the main text. The problem is filled with many local minima and therefore instead of learning good policies, the agent ends up memorizing nearby minima, which is easier to do when the initial value is fixed.

\section{Algorithmic error in NISQ era} 
\label{error-analysis}

Let's return to our original problem to begin our error analysis journey. We will first consider the error accumulated from measuring a single quantum state according to the setup described in sec.~(\ref{sec:quantum-processor}). From there, we will build up the entire training dataset, and examine various types of errors and mitigation strategies. We conclude that with our current NISQ setup, our measurements are too noisy to be used for gradient-based optimizers.

\subsection{Measuring a single quantum state}

For this analysis we will take a concrete data point:

\begin{align}
    \bar{x} &=
        \begin{bmatrix}
            0.0976 \\
            0.4304
        \end{bmatrix}
\end{align}

And we use a randomly initialized parameter $\bar{\theta}$:
\begin{equation}
\scalebox{0.65}{ % Adjust the scaling factor here
    $\bar{\theta}^T =
        \begin{bmatrix}
            0.1532 & 0.5374 & 2.3999 & -0.8025 & 0.3855 & 3.6122 & 1.2990 & 0.1235 & 1.3819 \\
            3.3182 & -2.4787 & 5.4758 & -1.4164 & 4.7989 & 0.5262 & 4.4542
        \end{bmatrix}$
    }
\end{equation}
%%%%%%%
% \begin{align*}
%     \bar{\theta}^T &=
%         \begin{bmatrix}
%             0.1532 & 0.5374 &
%             2.3999 &
%             -0.8025 &
%             0.3855 &
%             3.6122 &
%             1.299 &
%             0.1235 &
%             1.3819 &\\
%             3.3182 &
%             -2.4787 &
%             5.4758 &
%             -1.4164 &
%             4.7989 &
%             0.5262 &
%             4.4542
%         \end{bmatrix}
% \end{align*}
%%%%%%%%%%%%%%%
We will use Ansatz-2C with $L=4$. Thus we have:

\begin{align}
    \ket{\phi} &= U_3(\bar{\theta}, \bar{x}) U_2(\bar{\theta}, \bar{x}) U_1(\bar{\theta}, \bar{x}) U_0(\bar{\theta}, \bar{x}) \ket{0} \\
    &= R_z (\theta_{14} x_0 + \theta_{15} x_1) R_y (\theta_{12} x_0 + \theta_{13} x_1) \dots R_y (\theta_0 x_0 + \theta_1 x_1) \ket{0} \\
    &= R_z (0.2462) R_y (6.1721) \dots R_y (1.9684) \ket{0}
\end{align}

Following the above, we may compute $\ket{\phi} = \alpha \ket{0} + \beta \ket{1}$ and deduce that $\alpha^2 = 0.663$ and $\beta^2 = 0.337$.

\begin{figure}[h]
\begin{center}
    \includegraphics[width=0.8\linewidth]{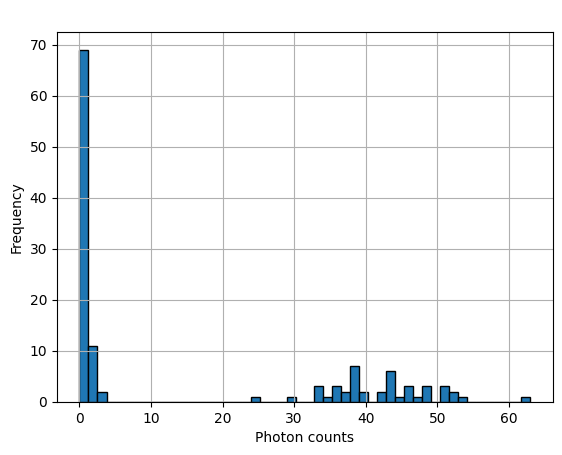} 
\end{center}
\caption{Histogram of 125 qubit projection shots. $\ket{0}$: photon counts between 0 and 12, and $\ket{1}$ photon counts greater than 12. Measured $\alpha^2=0.656$ and $\beta^2=0.344$.}
    \label{fig:single-state}
\end{figure}

We will now see what $\ket{\phi}$ looks like in our experimental setup in fig.~(\ref{fig:single-state}), where we performed $125$ repeated and independent measurements to estimate the value of $\beta^2$. Since we define $\ket{0}$ as the ground state and $\ket{1}$ as the excited state, the photon count will be higher ($>12$) when the qubit is in the excited state and lower ($<=12$) when its in the ground state. We can see that for this particular data point, our measurement value of $0.344$ is quite close to the theoretical value of 0.337.

\subsection{Measuring the training dataset}

In the previous section, we looked at how a single data point and parameter combination is measured using a qubit. We will now repeat this process for an entire training dataset.

\begin{figure}[hb]
\begin{center}
    \includegraphics[width=\linewidth]{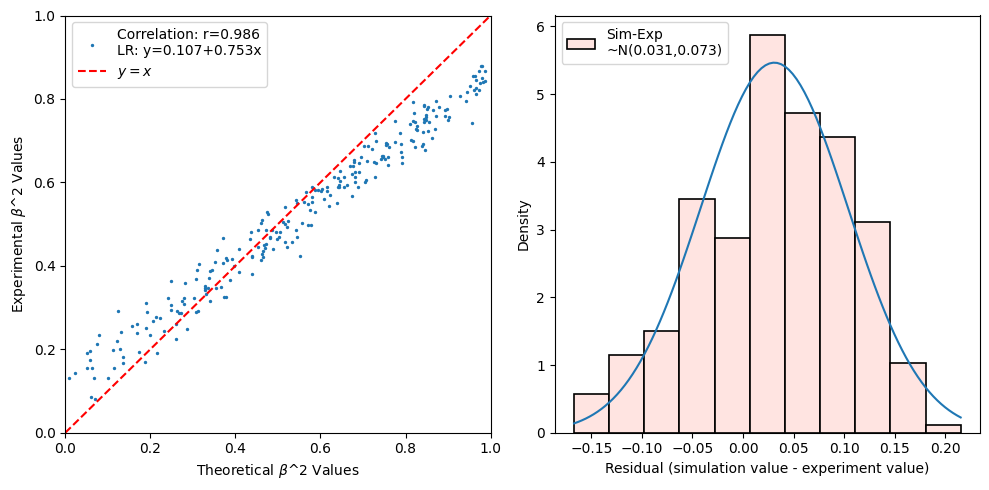} 
\end{center}
\caption{Left: repeated results from fig.~(\ref{fig:single-state}) over the entire training dataset and plot experimental against theoretical $\beta^2$ values. Post-application of inverse correlation matrix to experimental values is displayed in blue. Right: residuals (theoretical value - experimental value) of the results from the left. The histogram is fitted with a normal distribution and the y-axis is normalized using Gaussian density function.}
    \label{fig:training-dataset}
\end{figure}

We can learn a few things from fig.~( \ref{fig:training-dataset}). First, due to the natural decoherence of the qubit, the experimental measurements gradually tilts towards the $y=0.5$ line. In this particular case, the average amount of decoherence results in the regression line moving from $y=x$ to $y=0.106+0.759x$. This can be classified as a type of systematic error caused by the environment. Second, we see the dotted points scattered around the regression line. This is a result of random uncertainty from the experimental setup. This is a combination of systematic error from instruments and random error from environments. We will now attempt two separate error mitigation techniques to overcome the errors we identified above.
%%%%%%%%%%%%%%%%%%
\begin{figure}[ht]
    \centering
    \subfigure[\hspace{0mm} ]{
        \includegraphics[width=0.45\textwidth]{ga_varying_noise.png}
        \label{fig:ga_varying_noise}
        }
    \vspace{0 mm}
    \subfigure[\hspace{0mm} ]{
        \includegraphics[width=0.45\textwidth]{bfgs_0.005_varying_noise.png}
        \label{fig:bfgs_0.005_varying_noise}
        }
    \vspace{0 mm}
    \subfigure[\hspace{0mm} ]{
        \includegraphics[width=0.45\textwidth]{bfgs_0.05_varying_noise.png}
        \label{fig:bfgs_0.05_varying_noise}
        }
    \vspace{0 mm}
    \subfigure[\hspace{0mm} ]{
        \includegraphics[width=0.45\textwidth]{bfgs_0.5_varying_noise.png}
        \label{fig:bfgs_0.5_varying_noise}
        }
    \caption{Simulation of binary classification problem solved with a 4-layer data re-uploading architecture and varying Gaussian noise levels. All noises are assumed to have $0$ mean. (a) Genetic Algorithm solver with population size $50$, scattered crossover, exponentially decaying mutation rate, and sss parent selection scheme. (b) l-BFGS-b solver with stepsize (for gradient estimation) of $0.005$. (c) l-BFGS-b solver with stepsize (for gradient estimation) of $0.05$. (d) l-BFGS-b solver with stepsize (for gradient estimation) of $0.5$.
    }
    \label{fig:fig-extra1}
\end{figure}
%%%%%%%%%%%%%%%%%%%%%%
\subsection{Mitigation of systematic error}

A qubit in the real world experiences quantum decoherence overtime. This error can be mitigated post-measurement at the algorithmic level by defining the following matrix:

\begin{align}
    M &=
        \begin{bmatrix}
            P(\ket{0}|\ket{0}) & P(\ket{1}|\ket{0}) \\
            P(\ket{0}|\ket{1}) & P(\ket{1}|\ket{1})
        \end{bmatrix},
\end{align}

where $P(\ket{0}|\ket{1})$ represents the probability of measuring $\ket{0}$ when our theoretical prediction is $\ket{1}$. On our experimental setup, we prepare a sequence of dummy gates whose total gate time is the average of that of our training dataset. But those dummy gates produce end quantum states of either $\ket{0}$ or $\ket{1}$. Here is our measured matrix $M$:

\begin{align}
    M &=
        \begin{bmatrix}
            0.76 & 0.24 \\
            0.16 & 0.84
        \end{bmatrix}
\end{align}

We then apply $M^{-1}$ to our measurements in fig.~( \ref{fig:training-dataset}) and obtain the new error mitigated result shown in the same figure presented in blue.
%%%%%%%%%%%%%%%%%%
\begin{figure}[h]
\begin{center}
    \includegraphics[width=0.8\linewidth]{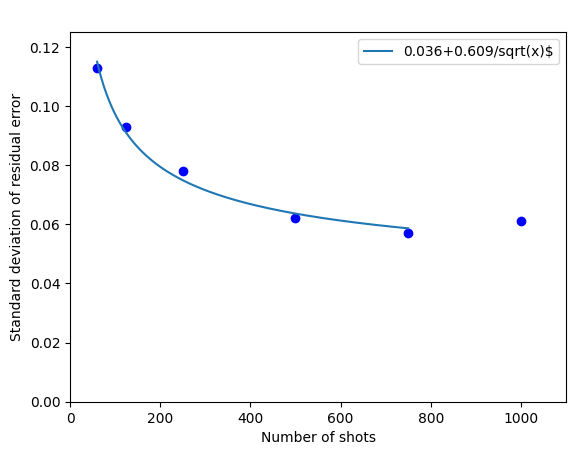} 
\end{center}
\caption{Standard deviation of residual errors (see example Figure~\ref{fig:training-dataset}) computed with different number of repetitions between 60 and 1000. The curve is fitted against the first 5 points.}
    \label{fig:training-dataset-noe}
\end{figure}
%%%%%%%%%%%%%
Comparing figures in~( \ref{fig:training-dataset}) shows that applying $M^{-1}$ to the experimental values helps mitigate systematic errors of the environment. The gradient of the regression line is closer to 1, and both the mean and standard deviation of the residual errors improved.
%%%%%
\begin{figure*}[t!]
\begin{center}
\subfigure[\hspace{0mm} ]{
\includegraphics[width=0.5\linewidth]{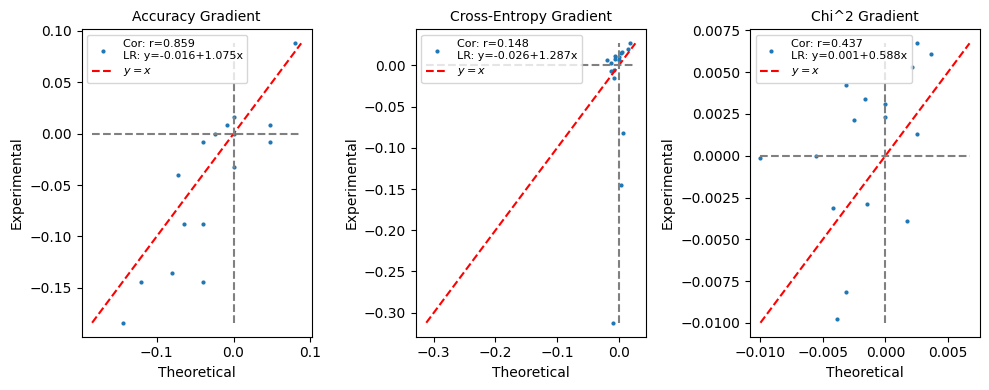}
      }
\subfigure[\hspace{0mm} ]{
\includegraphics[width=0.5\linewidth]{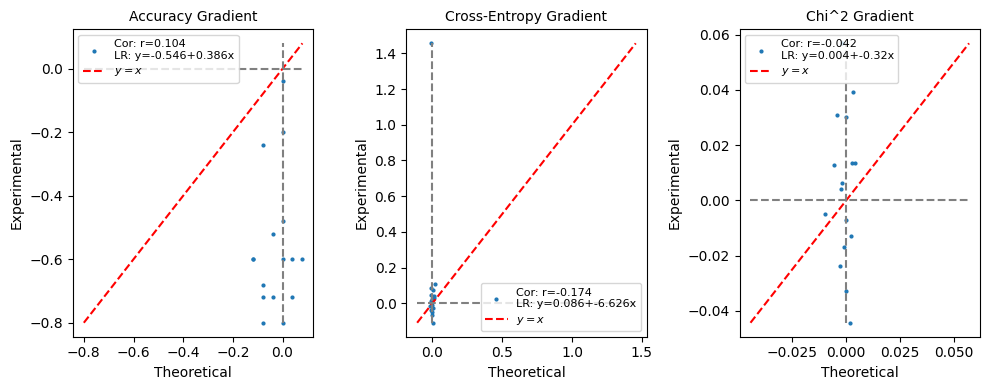}
      }
\end{center}      
\caption{Experimental vs theoretical gradients using three different metrics. (a) Gradient computed using step size of $0.5$. Dashed lines represent the x and y axes. Points in quadrants 2 and 4 represent experimental and theoretical gradients having opposite signs. (b) Gradient computed using step size of $0.1$.}
\label{fig:gradient-analysis}
\end{figure*}
%%%%%%%%
\subsection{Mitigation of random error}\label{sec:residual-error}
%%%%%
Figures in~(\ref{fig:training-dataset}) suggest that the random error from the experimental measurements are fairly normally distributed. If we follow this assumption, we expect our errors to decrease following $\frac{1}{\sqrt{N}}$, where N is the number of repetitions.

From fig.~(\ref{fig:training-dataset-noe}) we can see that the standard deviation of the residual error decreases proportional to $\frac{1}{\sqrt{N}}$. This trend stops beyond 750 repetition, where we start to observe random errors that do not scale with the number of repetitions. We refer to this as unknown or residual error. And this residual error of around $0.006$ represents the best experimental bound we may achieve under our current experimental setup.
%%%%%%%%%%%%
\subsection{Finite-difference based gradient optimizers}
In this sub-section, we examine the practicality of finite difference methods as our gradient-based optimizer. We see from fig.~(\ref{fig:gradient-analysis}) that our measured gradients using step sizes of $0.5$ and $0.1$ are all overridden by our noise, with the exception of the gradient of accuracy when step size is $0.5$. We proceeded with this setup in our experiment noting two caveats. Caveat one is that even under this setup, the sign of our gradients are sometimes on the wrong side, making it very difficult for the algorithm to improve. Caveat two is that step size of 0.5 is far too large for any gradient-based algorithms to work well, as is reflected in our experimental results.

In classical machine learning, we know that the step size should be chosen to be as small as possible for gradient-based optimizers to work well. Rigorously, the mathematical definition of gradient only exist as some small $\epsilon$ tends to zero. Luckily, for our particular type of quantum circuit, we may also employ the \textit{parameter-shift rule}~\cite{pennylaneParametershiftRules} to compute the gradients. We will explore this in the following section.
%%%
Additionally, we have conducted numerical experiments to demonstrate the robustness of the genetic algorithm (GA) over gradient-based methods. In Fig.~\ref{fig:fig-extra1}, we compare the performance of a typical GA solver against the l-BFGS-b solver, using various step sizes for finite difference gradient estimation. While the l-BFGS-b solver performs well in noise-free environments with smaller step sizes (see fig~\ref{fig:bfgs_0.005_varying_noise} and fig~\ref{fig:bfgs_0.05_varying_noise}), its performance deteriorates significantly with the introduction of noise. This behavior is expected, as the absolute values of the gradients are comparable in magnitude to the noise levels.
\\
In Fig.~\ref{fig:fig-extra}, we further analyze the performance of the GA solver and the l-BFGS-b solver under Gaussian noise with a standard deviation of $0.05$, consistent with our experimental setup. The noise-free results are presented alongside the noisy simulation results for comparison. Notably, the l-BFGS-b solver exhibits a significantly larger performance gap between noise-free and noisy conditions compared to the GA solver, highlighting the latter's robustness in noisy environments.
%%%%%%%%%%%%%%%
\subsection{Parameter-shift rule and limitation of gradient-based algorithms}
%%%
\begin{figure*}[t]
\begin{center}
    \includegraphics[width=0.9\linewidth]{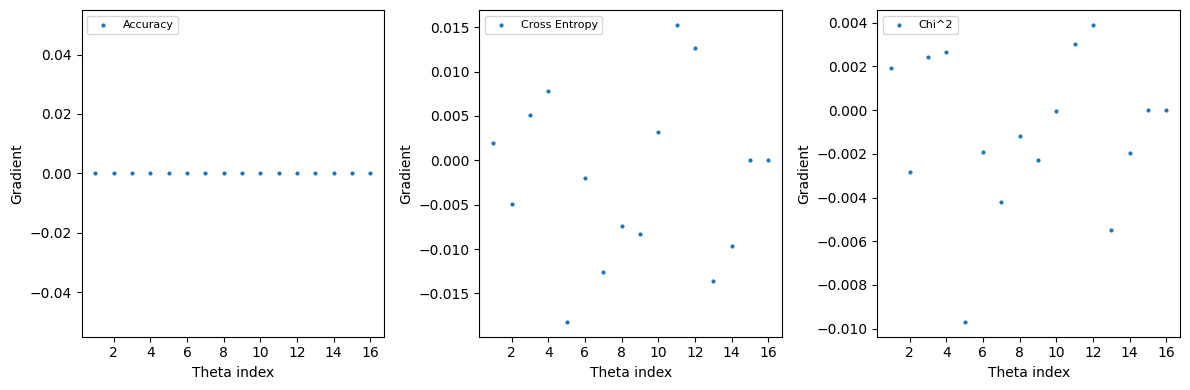} 
\end{center}
\caption{Partial derivative of Accuracy, Cross-entropy, Chi-squared values (see sec.~\ref{sec:cost-functions}) using finite difference with step size of $0.01$.}
    \label{fig:theoretical-gradients}
\end{figure*}
%%%%
Referring to fig.~(\ref{fig:theoretical-gradients}), we see the theoretical gradients of various optimizers using our given $\bar{\theta}$ of length $16$. The accuracy gradient is zero everywhere, matching our understanding that it is a \textit{sparse} signal (see sec.~(\ref{sec:cost-functions})). On the other hand, the gradients of cross-entropy and chi-squared values are non-zero but the magnitude of their values are extremely small - orders of magnitude smaller than the residual standard deviation from sec.~(\ref{sec:residual-error}). This makes all gradient-based methods undesirable candidates for optimization.\\
%%%%%%%%%%%%%%
%%%%%%%%%%%%%%%
% \bibliography{supplement}
% \end{document}
%%%%%%%%%%%%%%%%%%%%%%%%%%%%%%
\end{document}